\documentclass[12pt]{article}

\usepackage{newtxtext,newtxmath}

\usepackage{graphicx}

\usepackage[letterpaper,margin=1in]{geometry}

\renewenvironment{abstract}
	{\quotation}
	{\endquotation}

\date{}

\makeatletter
\renewcommand{\fnum@figure}{\textbf{Figure \thefigure}}
\renewcommand{\fnum@table}{\textbf{Table \thetable}}
\makeatother

\usepackage{scicite}

\usepackage{url}

\newcommand{\degree}{$^{\circ}$}
\newcommand{\angstrom}{\AA}
\newcommand{\micro}{$\mu$}

\newcommand{\SI}[2]{#1~#2}
\newcommand{\SIRef}[1]{SI \textsection\ref{#1}}

\def\scititle{
Stable electron-irradiated [1-$^{13}$C]alanine radicals for clinically viable metabolic imaging with Dynamic Nuclear Polarization
}
\title{\bfseries \boldmath \scititle\\ \small Short Title: Electron-irradiated radicals for DNP}

\author{
    Catriona~H.~E.~Rooney$^{1\dagger}$,
    Justin~Y.~C.~Lau$^{2\dagger}$,
    Esben~S.~S.~Hansen$^{3}$,\and
    Nichlas~Vous~Christensen$^{3}$,
    Duy~A.~Dang$^{3}$,
    Kristoffer~Petersson$^{4}$,\and
    Iain~Tullis$^{4}$,
    Borivoj~Vojnovic$^{4}$,
    Sean~Smart$^{4}$,\and
    William~Myers$^{5}$,
    Zoe~Richardson$^{6}$,\and
    Jarrod~Lewis$^{7}$,
    Brett~W.~C.~Kennedy$^{1}$,
    Alice~M.~Bowen$^{8}$,
    Lotte~Bonde~Bertelsen$^{3}$,\and
    Christoffer~Laustsen$^{3\ddagger}$,
    Damian~J.~Tyler$^{1, 9\ddagger}$, and 
    Jack~J.~Miller$^{3\ddagger}$\and
    \small$^{1}$Department of Physiology, Anatomy and Genetics, University of Oxford, UK\and
    \small$^{2}$GE HealthCare, Waukesha, WI, USA\and
    \small$^{3}$The MR Research Centre, Aarhus University, Aarhus, Denmark\and
    \small$^{4}$Department of Oncology, University of Oxford, UK\and
    \small$^{5}$Centre for Advanced ESR, Department of Chemistry, University of Oxford, UK\and
    \small$^{6}$Department of Physics, University of Oxford, UK\and
    \small$^{7}$Department of Material Science, University of Oxford, UK\and
    \small$^{8}${The National Research Facility for Electron Paramagnetic Resonance,}\and\small{The Photon Science Institute and Department of Chemistry, The University of Manchester}, UK\and
    \small$^{9}$OCMR, Cardiovascular Medicine, University of Oxford, UK\and
    \small$^\ast$Corresponding author. Email: jack.miller@physics.org;
    \small$^{\dagger,\ddagger}$These authors contributed equally to this work.
}

\begin{document} 

\maketitle

\clearpage
\begin{abstract} \bfseries \boldmath
Dissolution Dynamic Nuclear Polarisation (dDNP) increases the sensitivity of magnetic resonance experiments by $>10^4$-fold, permitting isotopically-labelled molecules to be transiently visible in MRI scans. dDNP requires a source of unpaired electrons in contact with labelled nuclei, cooled to $\sim$1K, and spin-pumped into a given state by microwaves. These electrons are usually chemical radicals, requiring removal by filtration prior to injection into humans. Alternative sources, such as UV irradiation, generate lower polarisation and require cryogenic transport. We present ultra-high-dose-rate electron irradiation as a novel alternative for generating non-persistent radicals in alanine/glycerol mixtures. These are stable for months at room temperature, quench spontaneously upon dissolution, are present in dose-dependent concentrations, and generate comparable nuclear polarisation to trityl radicals used clinically (20\%) through a novel mechanism. This process is inherently sterilising, permitting imaging of alanine metabolism \textit{in vivo}. As well as scientific novelty, this overcomes the biggest barrier to clinically translating dDNP. 
\end{abstract}

\textbf{Teaser}: 

High-energy electrons generate stable radicals for safer, more accessible metabolic imaging with dDNP through a novel mechanism. 

\noindent\textbf{\large\textsc{Main text}}
\noindent
\section{Introduction}

Hyperpolarised Magnetic Resonance Imaging (HP MRI) is a molecular imaging technique that is predominantly used to monitor metabolic fluxes in real-time by transiently making a labelled molecular probe visible to magnetic resonance experiments such as MRI. Over the past two decades, its pre-clinical use has spanned multiple fields of basic scientific and medical research, with 2013 marking the start of its validation for various clinical applications, including prostate cancer\cite{Nelson2013}, traumatic brain injury\cite{Hackett2020}, and ischaemic heart disease,\cite{Apps2021} amongst many others.\cite{Larson2024} To perform HP MRI, a hyperpolariser is required -- a device that transiently enhances the MR signal of stable labelled isotopes by several orders of magnitude. One method to achieve this is through dissolution Dynamic Nuclear Polarisation (dDNP), which employs cryogenic temperatures, high magnetic field strengths ($>3$\,T), and microwave irradiation. A typical sample for dDNP comprises a metabolic substrate mixed with a chemical radical species and optionally also a glassing agent such as glycerol. At low ($\sim1$ K) temperature, the unpaired electron spins present in the radical become thermally polarised. When irradiated with microwaves, a population inversion of these electronic spins can be created, and they are driven to relax, primarily through dipolar interactions, and ultimately transfer their polarisation to nearby nuclei. The net process behind DNP therefore is that nuclear polarisation above that at thermal equilibrium is created. Once the desired nuclear signal enhancement is achieved, a heated and pressurised solvent is introduced to rapidly melt the solid-state sample and convert it into a form suitable for injection into a living system. Following injection, the enzymatic conversion of the hyperpolarised substrate into other metabolic intermediates can be temporally and spatially monitored \textit{in vivo} through magnetic resonance experiments such as MRI. 

Hyperpolarised pyruvate has become the most widely used substrate in HP MRI due to its molecular utility and importance in biochemistry, and the high degree of polarisation ($>50\%$) that can be obtained with dDNP.\cite{Ardenkjaer-Larsen2019} However, in clinical settings, its preparation is complicated by the relative toxicity of the chemical radicals used. This requires either (1) chemical engineering to produce a soluble trityl radical that can be filtered out -- the approach taken in all clinical trials to date, as the LD\textsubscript{50} of trityl radicals is approximately 8~mmol/kg\cite{Serda2016} and they are present in samples at concentrations above this -- or (2) photon irradiation to generate transient radicals that serve as the electron source necessary for dDNP and quench rapidly upon dissolution.\cite{Capozzi2017} The key advantages of the latter approach are the elimination of the filtration step, enabling faster and more efficient use of the sample, and the extended duration of the enhanced MR signal (i.e.~a lengthened $T_1$ relaxation time), as it avoids the presence of unpaired electrons that act as relaxation centres following DNP\cite{Pinon2020}. To date, both ultraviolet\cite{Eichhorn2013} and gamma irradiation\cite{Giannoulis2024} have shown potential. 

However, regardless of whether the electron source is chemical or photon-based, there are substantial technical limitations on the ability of clinical sites to ship and receive compounds for hyperpolarisation with the radicals \emph{in situ}. For instance, the chemical instability of the trityl radical --- the most commonly used chemical radical --- renders centralised production of pre-filled fluid paths for global distribution infeasible;  fluid paths are transported typically at $195$\,K. In contrast, while low-energy gamma-generated radicals are reported to be stable at room temperature for several months to produce a single sample, irradiation can take days to months, and reported nuclear enhancement factors are on the order of $10^2$ and not the  $10^5$ obtained with chemical radicals.\cite{Jacobs2022} DNP with UV-generated radicals is only possible if samples are optically transparent (e.g.~pyruvic acid) and require low-temperature (cryogenic) transportation\cite{Capozzi2017}, which incurs a significant expense and is not routinely performed as precise control of $B$ and $T$ is required over many orders of magnitude.\cite{Capozzi2025} Ultimately, sites conducting human dDNP experiments generally require local access to a pharmacy equipped with sterile compounding facilities to comply with the high degree of sterility assurance required. This is complex from a regulatory view,\cite{Larson2024} and is considered a significant financial burden that may ultimately limit the accessibility of the technique. 

Here, we present an alternative approach for generating sterilised samples for clinical HP MRI that are suitable for convenient transport and centralised manufacture, potentially offering an alternative for sites that do not have access to a pharmacy. Using an ultra-high dose rate 6 MeV electron linear accelerator (linac), we demonstrate the creation of endogenous radicals within polycrystalline alanine, the concentration of which can be easily tuned by adjusting properties of the e-beam. Alanine is commonly used for radiation dosimetry, and the radicals generated upon irradiation are known to be stable for many years at room temperature if stored anhydrously. To improve thermal conductivity and permit efficient spin diffusion between polarising centres, irradiated polycrystalline alanine was dispersed in glycerol, resulting in radical lifetimes that were stable for several weeks when stored at room temperature with silica desiccant gel. Upon dissolution, radicals were quenched, yielding a non-toxic, directly injectable substrate. At 6.7 T, we were able to obtain nuclear polarisation levels of approximately 20\% and discovered that the polarisation transfer is not simply explained by the solid effect or models of thermal mixing with the known stable alanine radical $g$-factor and hyperfine couplings at this field strength. This is unlike similar analytic models of DNP with pyruvate and disperse chemical free radicals such as trityl\cite{Johannesson2019}. Crucially, the hyperpolarised \textsuperscript{13}C label on alanine was successfully transferred to pyruvate and lactate after injection \textit{in vivo}, demonstrating the potential of this method for metabolic imaging.
\section{Results}

\subsection{Irradiation of Samples}

\begin{figure}
    \centering
    \includegraphics[width=0.8\linewidth]{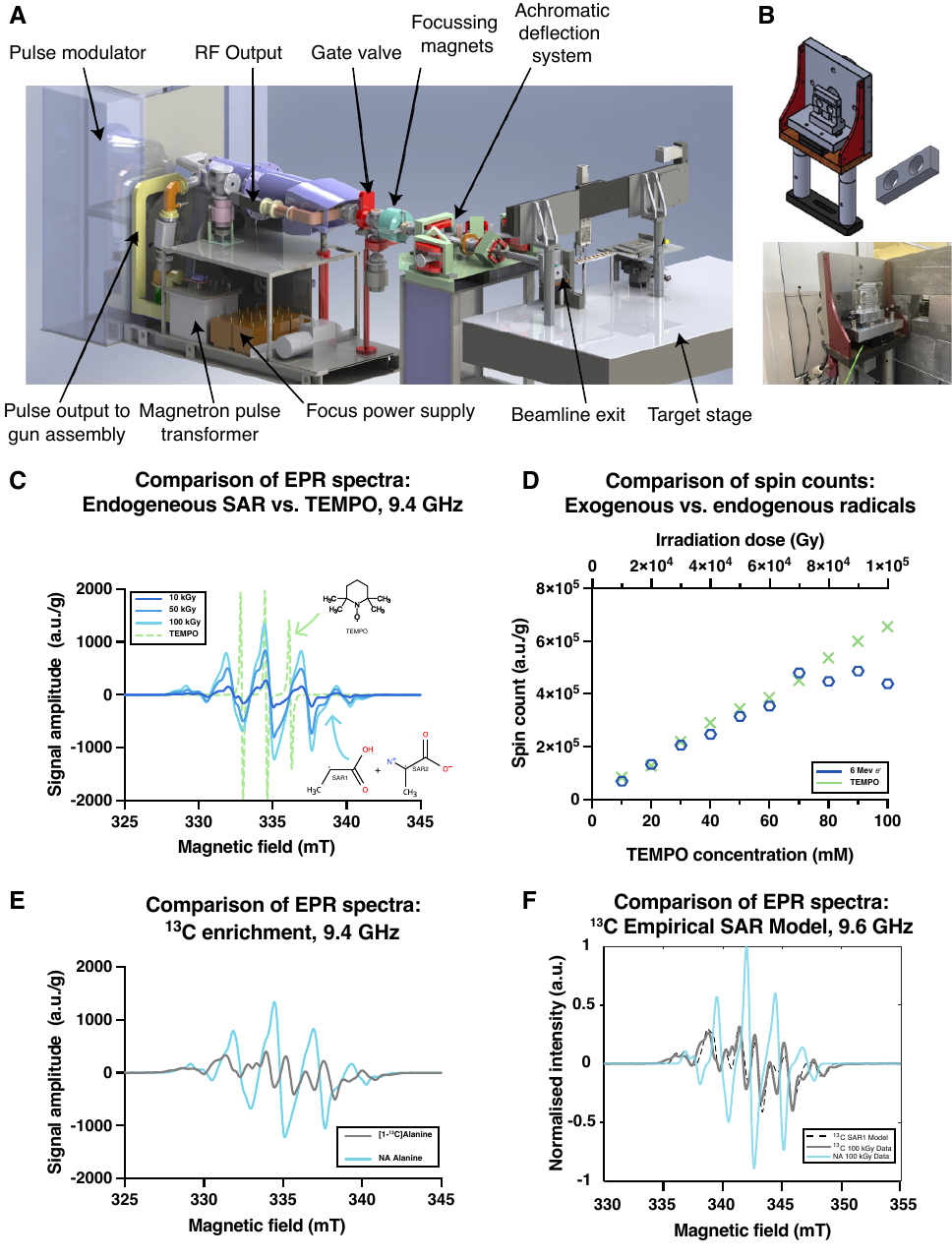}
    \caption{Radical generation. (\textbf{A}) An ultra-high dose rate optimised in-house developed linear accelerator was used to bombard (\textbf{B}) a target containing polycrystalline [1-\textsuperscript{13}C]alanine with \SI{6}{MeV} electrons. (\textbf{C}) These generate Stable Alanine Radicals (SAR) detectable by EPR in a dose-dependent fashion compared to (\textbf{D}) TEMPO as a concentration standard. (\textbf{E}) The \textsuperscript{13}C-hyperfine interaction is strong for the SARs, significantly altering the EPR spectrum compared to natural abundance (NA) alanine, which we quantified (\textbf{F}) via simulations in EasySpin.}
    \label{fig:linac-and-radicals}
\end{figure}

Dry polycrystalline natural-abundance (NA) and [1-\textsuperscript{13}C]L-alanine powders were irradiated using a FLASH-optimised\cite{Vozenin2022} in-house developed 6 MeV (nominal energy) electron linear accelerator.\cite{Berne2021} This specialised accelerator (Fig.~\ref{fig:linac-and-radicals}\textbf{A}) was based on a reconfigured Elekta SL75-5 travelling-wave waveguide with an S-band RF magnetron source (\SI{2.89}{GHz}) and delivers ultra-high dose rate horizontal \SI{6}{MeV} electron beam \SI{3.5}{\micro s} macropulses, with adjustable pulse repetition rates (25-\SI{300}{Hz}) through precise control of RF frequency and injection parameters. \cite{Berne2021} Unlike conventional medical linacs that employ \SI{90}{\degree} bending magnets, our system delivered electrons in a straight-through beam path with specialised quadrupole focusing and advanced beam monitoring systems,\cite{Berne2021a, Vojnovic2023} allowing exquisite control of dose-per-pulse across multiple orders of magnitude, and was uniquely able to irradiate alanine samples, provided as a powder sample (Fig.~\ref{fig:linac-and-radicals}\textbf{B}), with up to \SI{100}{kGy} in seconds. Further details are provided in the \SIRef{SI:irradiation}.

\begin{center}
    {\small{\emph{[Figure 1 approximately here]}}}
\end{center}

We undertook EPR spectroscopy to quantify radical species generated by electron irradiation. As radiation damage in dry, natural-abundance (NA) alanine is known to produce stable alanine radicals species (SARs) that have been extensively studied, both via EPR and theoretically,\cite{Heydari2002, Janbazi2018, Jastad2017, Malinen2003, Pauwels2014, Sagstuen1997a,Matsuki1982,Miyagawa1960} we compared experimental EPR spectra to that of TEMPO ((2,2,6,6-Tetramethylpiperidin-1-yl)oxyl), an aminoxyl radical used for DNP, and additionally quantitatively verified their dose/response curve. 
Endogeneous alanine radicals had a broader linewidth than TEMPO (Fig.~\ref{fig:linac-and-radicals}\textbf{C}) and, as expected an increase in the irradiation dose led to a greater spin count, saturating at $\sim$\SI{70}{kGy} (Fig.~\ref{fig:linac-and-radicals}\textbf{D}). We found that [1-\textsuperscript{13}C]alanine had a drastically different EPR spectra compared to NA alanine, reflecting a large hyperfine interaction (not altering the corresponding spin count; \SIRef{SI:SpinCount}). The $g$-tensor and proton/nitrogen hyperfine interaction constants $A$ for NA alanine have previously been well documented;\cite{Kuroda1982, Miyagawa1960, Pauwels2014, Sagstuen1997} however, no such terms for [1-\textsuperscript{13}C]alanine have been measured. We determined using a constrained Bayesian optimisation routine\cite{Martinez-Cantin2014} the values of $A$ and $g$ for the predominant SAR species, known as R1 (Fig.~\ref{fig:linac-and-radicals}\textbf{F}; \SIRef{SI:AlanineSAROptimisation}); the principal values determined for $A$ were $[-1.80266,\, -15.8427,\, 64.6933]$~MHz with Euler angles of $[1.50914,\, 1.30749,\, 0.761318]$~rad in the previously reported molecular frame\cite{Pauwels2014}.

\subsection{DNP via Irradiation-Induced Radicals}

\begin{figure}[htbp]
\centering
\includegraphics[width=\textwidth]{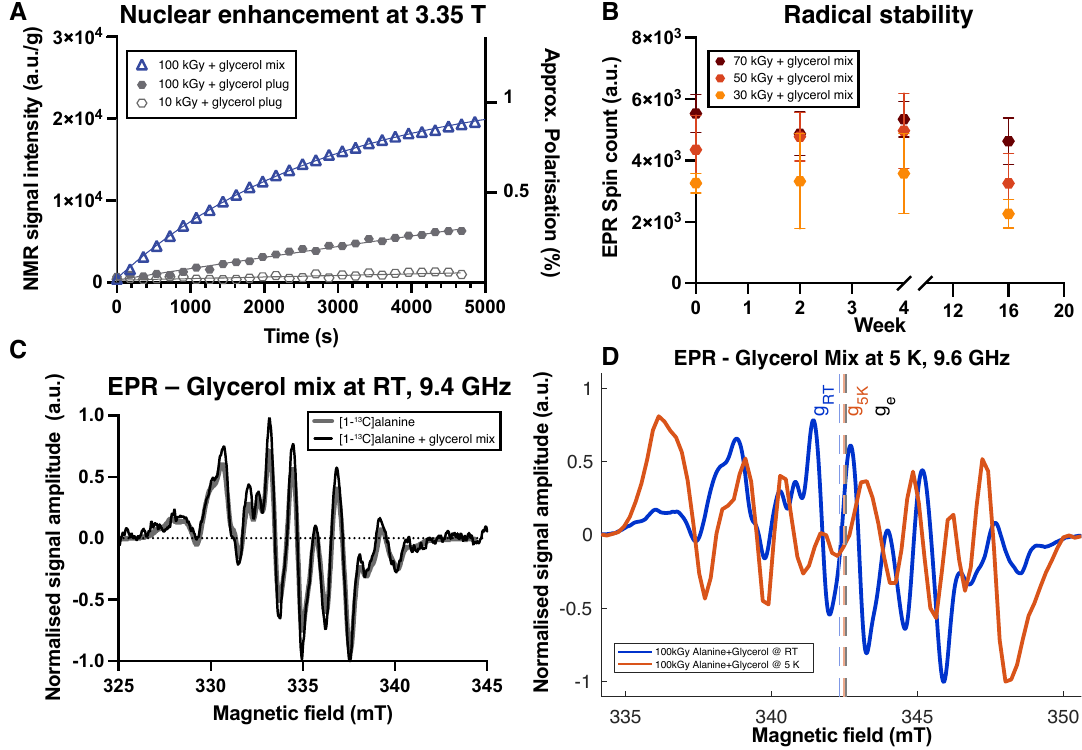}%
\caption{(\textbf{A}) DNP build-up profiles of different high dose rate irradiated samples at \SI{3.35}{T}; we found that dispersing irradiated polycrystalline alanine powder in glycerol produced substantially higher polarisation. (\textbf{B}) Longitudinal measurements of the spin counts showed that there was no significant effect on the endogenous radical concentration with increased duration of storage in anhydrous glycerol up to 4 months, in sharp contrast to pyruvic acid/trityl radical mixtures which quench within hours (mean $\pm$ SD of three technical replicates). (\textbf{C}) Alanine/glycerol mixtures did not significantly alter acquired EPR data, but at (\textbf{D}) \SI{5}{K} the electronic environment altered considerably }
\label{fig:DNP}
\end{figure}

We first tested the feasibility of utilising the irradiation-induced radicals for DNP by exploring their polarisation characteristics and stability during sample storage. We investigated both dry powder mechanically secured in place with a separate snap-frozen ``plug'' of anhydrous glycerol and an alanine and glycerol mixture. We found that limiting nuclear polarisation increased with increasing radical concentration, but that significantly greater nuclear enhancement at 3.35~T was obtained with alanine crystallites dispersed in the glycerol matrix (Fig.~\ref{fig:DNP}\textbf{A}). These mixtures were stable for weeks independent of radiation dose, as quantified by EPR (Fig.~\ref{fig:DNP}\textbf{B}), with no significant change in concentration vs time. The addition of glycerol did not change the electronic environment at room temperature (Fig.~\ref{fig:DNP}\textbf{C}), but did at 5~K (Fig.~\ref{fig:DNP}\textbf{D}) with substantially altered hyperfine interaction constants and a shift in the mean isotropic electronic $g$-value (from $g=2.004100$ to $g=2.00169$, shown integrated in Fig.~\ref{fig:thermalmixing}) that we were not able to model. 
\begin{center}
    {\small{\emph{[Figure 2 approximately here]}}}
\end{center}

\begin{figure}[htbp]
\centering
\includegraphics[width=0.65\textwidth]{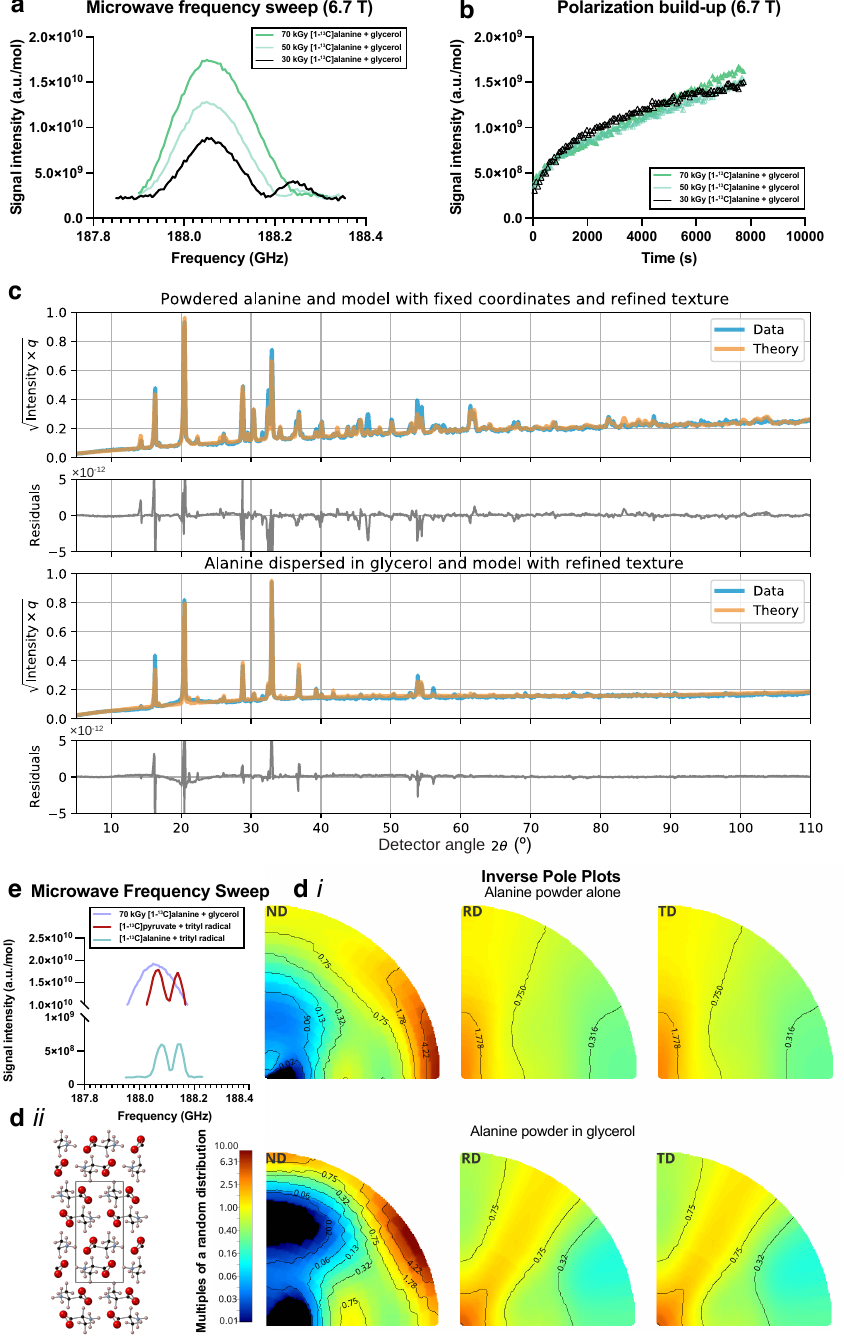}
\caption{(\textbf{A}) The frequency sweep profile of DNP with alanine/glycerol mixtures showed qualitative differences with increasing radiation dose, which we hypothesise is due to cooperativity. (\textbf{B}) Polarisation build-up as a function of time showed profound biexponential behaviour under these conditions, distinct from narrow linewidth radicals such as trityl. (\textbf{C}, \textbf{D}\emph{i}) X-ray diffractometery measurements of both samples were able to demonstrate that polycrystalline alanine remained dispersed (not dissolved) in glycerol and was well described by (\textbf{D}\emph{ii}) existing crystal structures with slight refinement but had profound crystallographic texture, indicative of an ordered crystalline arrangement within the glycerol matrix. (\textbf{E}) The efficiency of DNP under these conditions is surprising, comparable to narrow linewidth trityl radicals.}
\label{fig:microwave}
\end{figure}

Given the comparatively wide EPR linewidth of alanine radicals at \SI{3.35}{T}, we elected to undertake DNP at \SI{6.7}{T}. Under these conditions the relatively small degree of $g$ anisotropy would be expected to scale proportional to $B_0$ whereas the strong hyperfine couplings are field independent (depending primarily on the Fermi contact mechanism). We therefore expected to a first approximation an improvement in DNP efficiency through an analytic model of the thermal mixing process of polarisation growth based on the Provotorov equations,\cite{Capozzi2019} as the effective width of the EPR lineshape $D$ would be expected to decrease. We additionally explored radiation doses (\SI{30}{kGy}, \SI{50}{kGy}, and \SI{70}{kGy}) to reflect the concentrations of broad line width exogenous chemical radical species typically used for DNP (i.e., 30~mM, 50~mM, and 70~mM). To our surprise, we did not observe the characteristic ``bimodal'' frequency sweep pattern associated with either the well-resolved solid effect or thermal mixing mechanisms, the latter of which is expected to describe DNP with nitroxide or trityl radicals and produces maxima approximately at the sum or difference of the nuclear and electron Larmor frequencies ($\omega_e \pm \omega_N$). Instead we observed an asymmetric ``hump'' at higher frequency to the major lobe of the curve that decreased with increasing radiation dose (i.e. approximately at $\omega_e$ alone) as illustrated in Fig.~\ref{fig:microwave}\textbf{A}.
At lower doses (\SI{30}{kGy}), we observe a bimodal microwave profile with a prominent second lobe that progressively diminishes as the radiation dose increases to \SI{50}{kGy}, and ultimately disappears entirely at \SI{70}{kGy}. This systematic change suggests a transition in the underlying polarisation transfer mechanism that correlates directly with radical concentration. A biexponential build-up profile was also observed (Fig.~\ref{fig:microwave}\textbf{B}). 

To understand this difference upon the addition of glycerol, we undertook X-ray diffraction. Data were obtained from natural abundance polycrystalline alanine and polycrystalline alanine dispersed in glycerol; a spherical harmonic Rietveld crystallographic texture refinement\cite{McCusker1999} aiming to quantify any degree of order in the sample was undertaken with MAUD.\cite{Lutterotti1999} Using a neutron-diffraction crystal structure obtained on the SXD instrument\cite{Keen2006} at the ISIS neutron source,\cite{Wilson2005} we demonstrated strong partial ordering of the sample (Fig.~\ref{fig:microwave}\textbf{C}) that changed (and strengthened) upon addition of glycerol (Fig.~\ref{fig:microwave}\textbf{D}). As is common in texture analysis, we express orientation in units of multiples of a random distribution (\emph{mrd}), where a sample without preferred orientation is one uniformly; alanine and glycerol mixtures had orientation preferences occurring $>\SI{10}{\emph{mrd}}$ and are partially ordered. This was consistent with \emph{ab initio} liquid-state molecular dynamic simulations performed on two semi-infinite glycerol/alanine domains in LAAMPs\cite{LAMMPS} with EMC\cite{InTVeld2003} that predicted the immiscibility of the two substances (\SIRef{SI:MolecularDynamics}). 

\begin{center}
    {\small{\emph{[Figure 3 approximately here]}}}
\end{center}

This (Fig.~\ref{fig:microwave}\textbf{A}) microwave sweep profile is distinct from others previously reported, and the mechanistic processes by which DNP occurs appear as efficient as trityl radicals used in clinical research. We therefore undertook control DNP experiments comprising substrates polarised via exogenous chemical radical trityl species, namely \textsuperscript{13}C-pyruvate with \SI{30}{mM} AH111501 and \textsuperscript{13}C-alanine with \SI{38}{mM} OX063 Trityl (Fig.~\ref{fig:microwave}\textbf{E}).  We observed that the greater the radical concentration, the longer the slow component of the build-up time and the shorter the fast component (\SIRef{SI:builduptimes}). 

\begin{figure}[htpb]
    \centering
    \includegraphics[width=\linewidth]{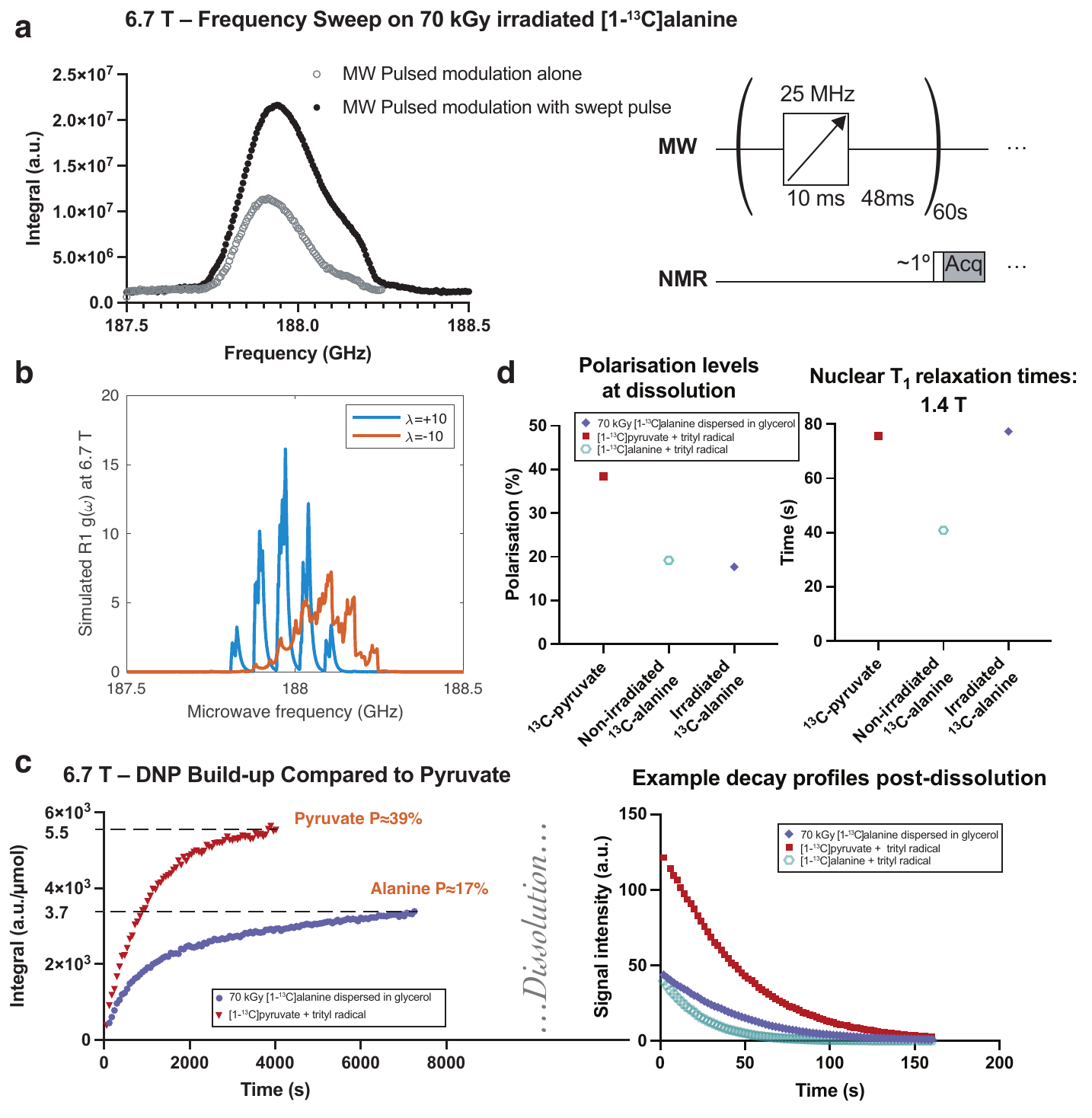}
    \caption{(\textbf{A}) At 6.7 T, a unimodal, always-positive frequency sweep curve was observed. Pulsed modulation with a swept-frequency microwave pulse as illustrated obtained a significant increase in nuclear polarisation. (\textbf{B}) The molecular orientation of the alanine radical matters significantly for the spectral density function $g(\omega)$ as defined by Wenckebach,\cite{Wenckebach2017a}  simulated here using parameters from powder \textsuperscript{13}C alanine EPR data and assuming a partial ordering parameter $\lambda$. (\textbf{C}) The resultant limiting polarisation obtained was high, and comparable to the use of trityl radicals with alanine. (\textbf{D}) After dissolution, spontaneous quenching of endogeneous radicals formed by electron irradiation resulted in a lengthened $T_1$ (\SI{77}{s}) compared to that with trityl (\SI{40}{s}) at \SI{1.4}{T}.}
    \label{fig:freq-sweep-and-dissolution}
\end{figure}

Given the wide EPR lineshape of the stable alanine radical, we investigated the use of a frequency-swept microwave pulse sequence for DNP. This resulted in an increased polarisation transfer and increased enhancement, within a factor of 2.5 of the ``gold standard'' pyruvate with trityl radical (Fig.~\ref{fig:freq-sweep-and-dissolution}\textbf{A}). In an attempt to understand the mechanism behind this observation, we simulated the fitted EPR lineshape under DNP conditions. We found that a polycrystalline distribution of R1 radicals have distinct spectral density functions $g(\omega)$ depending on preferred orientation, quantified by the order parameter $\lambda$ where the angle $\alpha$ between the molecular $z$-axis and $B_0$ is distributed by $p(\alpha)=\exp\left(-\lambda \left(3\cos^2 \alpha -1\right)/2\right)$, with the orientation preference found by X-ray diffraction to correspond to a combination of these parameters (Fig.~\ref{fig:freq-sweep-and-dissolution}\textbf{B}). A detailed quantum-mechanical simulation of this pulse sequence together with the known radical parameters was then undertaken in the advanced magnetic resonance simulation environment Spinach.\cite{Hogben2011} 
Despite including more than 50 \textsuperscript{13}C spins representing an alanine supercell with detailed measurements of its proton and carbon chemical shift tensors,\cite{Kruse2012, Kuroda1982}  simulating the entirety of the minute-long experiment at high temporal resolution (compared to the electron Larmor frequency) and utilising the Arcus-C supercomputing cluster,\cite{richards_2015_22558} we were unable to reproduce this experimental result and instead predicted a bimodal curve qualitatively similar to that expected for thermal mixing (\SIRef{si:spinach}). Previously published analytic models of DNP via thermal mixing and the cross effect\cite{Capozzi2019, Wenckebach2017a, Wenckebach2019a} likewise produce a predicted bimodal curve with either predicted analytic or experimental EPR data (\SIRef{si:analyticmodels}). A pseudopotential plane-wave simulation in QuantumESPRESSO\cite{Giannozzi2009} with our structures predicts that alanine crystals are a direct-gap semiconductor with a large (\SI{5}{eV}) bandgap (\SIRef{si:QE}, experimentally validated\cite{Suresh2020}); were the effect of electron irradiation sufficient to dope spins into the conduction band, we would expect to see a well-resolved solid effect mechanism and narrow EPR data accordingly. 
This, again, did not match the data. We hypothesise that the density of unpaired electrons is insufficient to form a delocalised conduction band but rather creates localised mid-gap states. This intermediate regime, where radicals are neither fully isolated molecules (as often assumed in molecular DFT calculations used for alanine radicals\cite{Pauwels2014}) nor fully delocalised carriers (as in classical solid effect DNP systems such as paramagnetic salts\cite{Abragam1978}), may explain why neither approach fully captures our experimental observations. The emergence of a complex and strongly orientation-dependent EPR lineshape suggests that these localised states retain molecular character in diffuse partially ordered systems while experiencing significant inter-radical coupling mediated by the alanine crystal lattice and glycerol matrix, necessitating models that bridge molecular and solid-state treatments that we were unable to provide.

Nevertheless, encouraged by the feasibility of polarising [1-\textsuperscript{13}C]alanine via irradiation-induced radicals, our focus turned to exploiting this for biomedical imaging. Empirically optimised DNP schemes compatible with our hardware produced an estimated solid state polarisation of approximately 20\% (Fig.~\ref{fig:freq-sweep-and-dissolution}\textbf{C}), similar to that obtained with a separate sample of [1-\textsuperscript{13}C]alanine and OX063 trityl radical, both of which are comparable to the ``gold standard'' used clinically of pyruvate with the trityl radical AH111501, which polarised to approximately 40\%. After dissolution, the immediate quenching in aqueous media of endogenous alanine radicals resulted in an increase in the apparent liquid-state $T_1$ at \SI{1.4}{T} (Fig.~\ref{fig:freq-sweep-and-dissolution}\textbf{D}), as may be expected from the removal of paramagnetic relaxation centres.\cite{Eichhorn2013}

\begin{center}
    {\small{\emph{[Figure 4 approximately here]}}}
\end{center}

\subsection{\emph{In Vivo} Application as a Metabolic Tracer}

\begin{figure}[htbp]
\centering
\includegraphics[width=\textwidth]{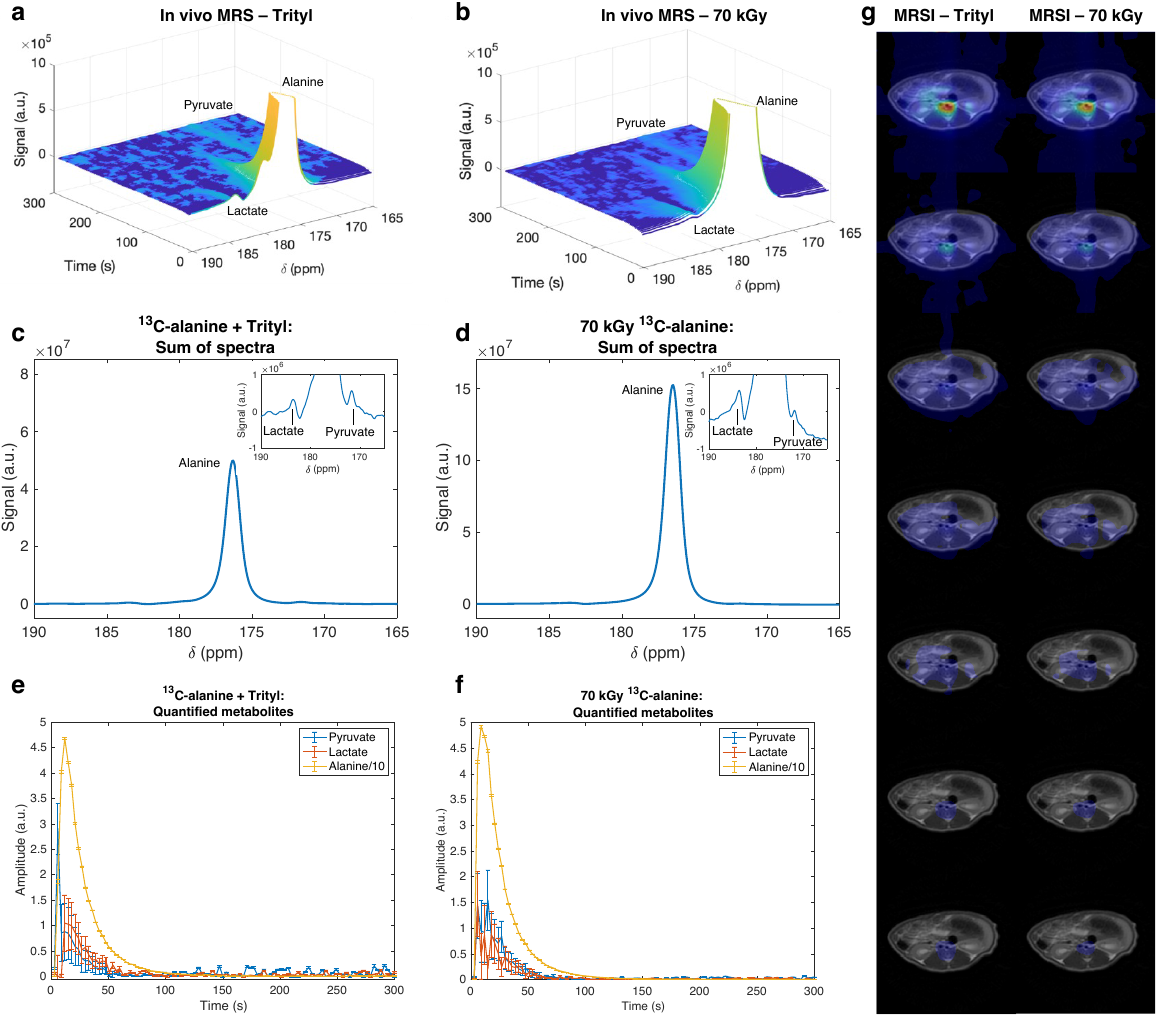}
\caption{\textbf{(A, B)} Stacked and \textbf{(C, D)} summed spectra from slice-selective spectroscopic acquisitions following separate tail-vein injections of [1-\textsuperscript{13}C]alanine polarised via \SI{38}{mM} OX063 trityl radical and hyperpolarised irradiated \textsuperscript{13}C-alanine. (\textbf{C, D}) Inset: a magnification of the spectra presented in the panels to highlight the intensities and positions of the downstream metabolites, whose chemical shift values match those reported in the literature for \textsuperscript{13}C-lactate and \textsuperscript{13}C-pyruvate. (\textbf{E, F}) After quantification by AMARES, metabolite dynamics were comparable between the two methods of preparation. (\textbf{G}) MR Spectroscopic Imaging (MRSI) images obtained in a separate rat were likewise near identical, validating the use of the technique for \emph{in vivo} imaging of redox potential, alanine images of which are shown concatenated over time.}
\label{fig:in_vivo}
\end{figure}

To test the optimised irradiated alanine preparation and DNP protocol, two proof of concept \textit{in vivo} studies were conducted in rats. First, slice-selective magnetic resonance spectroscopy acquisitions were performed following separate tail-vein injections of [1-\textsuperscript{13}C]alanine polarized with trityl radicals as described previously\cite{Nielsen2020} or endogenous radicals generated via 70 kGy electron-beam irradiation. Temporal spectroscopic analysis demonstrated higher initial signal magnitude from the irradiated sample relative to the trityl-based preparation (Fig.~\ref{fig:in_vivo}\textbf{A}, \textbf{B}). Secondary resonances were observed flanking the primary [1-\textsuperscript{13}C]alanine signal in both preparations. Spectral summation was performed to enhance the signal-to-noise ratio (SNR) and facilitate metabolite identification (Fig.~\ref{fig:in_vivo}\textbf{C}, \textbf{D}). The chemical shifts of these secondary resonances (\SI{183}{ppm} and \SI{171}{ppm}) corresponded precisely with the literature-reported values for [1-\textsuperscript{13}C]lactate and [1-\textsuperscript{13}C]pyruvate, respectively\cite{Hansen2022}. Subsequent metabolite quantification via the AMARES algorithm revealed comparable metabolic dynamics between both preparation methods (Fig.~\ref{fig:in_vivo}\textbf{E}, \textbf{F}) and spectral SNR was sufficient to quantify the downstream metabolites expected \emph{in vivo}. No additional labelled peaks were observed; the products of the dissolution were identical to that observed with trityl radical. The dissolution was `clean', and a final concentration of approximately 100~mM dissolved alanine was obtained. 

\begin{center}
    {\small{\emph{[Figure 5 approximately here]}}}
\end{center}

Secondly, comparative 2D magnetic resonance spectroscopic imaging (MRSI) was performed in a separate animal using identical hyperpolarised [1-\textsuperscript{13}C]alanine preparations. MRSI data were acquired, together with $T_2$-weighted proton images acquired in the axial plane for anatomical localisation. Spatiotemporal analysis of the spectroscopic data demonstrated concordant biodistribution patterns between the trityl-based and irradiation-based preparations (Fig.~\ref{fig:in_vivo}g), with predominant accumulation in the renal parenchyma as alanine perfuses the kidneys, consistent with previous renal reports and enabling the technique to quantify redox potential,\cite{Nielsen2020} and confirming equivalent \emph{in vivo} behaviour between preparation methods. 

\section{Discussion}

The clinical feasibility and utility of metabolic imaging with hyperpolarised MRI and dDNP has been consistently demonstrated since its inception 22 years ago \cite{Ardenkjaer-Larsen2003}, with the transformative ability to spatially map metabolism of huge interest in many disparate diseases including cancer,\cite{Nelson2013} heart disease,\cite{Apps2018} and diabetes\cite{Rider2020}. With more than 1000 human patients scanned and multi-centre trials soon to commence\cite{Larson2024a} it is a physical technique that has opened new avenues for research and the diagnosis and management of disease. However, for its integration within routine clinical workflows to be fully realised, significant improvements in its cost, accessibility and practicality are essential. Beyond the substantial capital expenditure for dDNP hardware (typically exceeding \$2 million), institutions using dDNP face prohibitive ongoing operational costs. These include maintaining dedicated clean-room pharmacy facilities with specialised personnel for the sterile preparation of per-patient disposable fluid paths for dDNP, implementing expensive quality control measures, and adhering to stringent regulatory frameworks for pharmaceutical preparation. Each hyperpolarised examination currently requires the labour-intensive preparation of sterile fluid paths,\cite{Ardenkjaer-Larsen2011} their  filling with sterile stable-isotope labelled metabolites together with dissolution and neutralisation media (both manufactured according to sterile guidelines that vary regionally\cite{Larson2024a}) and a trityl radical that provides the unpaired electron the technique requires. The synthesis of trityl radicals is not trivial, typically requiring $\sim$13 separate organic synthetic steps, leading to a cost $>$\$10k/g.\cite{Serda2016} Moreover, the required subsequent filtration steps to remove the radical add further complexity to the design of the fluid path. These are all individual processes that substantially increase per-patient costs and logistical complexity and restrict the availability of a technique that has the ability to directly measure central metabolic reactions key to all life. Furthermore, the limited shelf-life of the radical in prepared samples (approximately 24 hours at room temperature for trityl\cite{Ardenkjaer-Larsen1998}) necessitates precise coordination between pharmacy operations and clinical scheduling, and prohibits the transport of dDNP samples nationally or internationally unless extraordinary steps are taken such as transporting samples cryogenically under defined (and varying) temperatures and magnetic fields.\cite{Ji2017, Kiryutin2019a,Capozzi2025} As well as restraining human experiments, these large costs further constrain dDNP availability in non-human biological research: although of utility studying species such as snakes\cite{Hansen2022} and veterinary canine patients,\cite{Gutte2015} trityl radicals are sufficiently expensive as to be repurified from animal urine after use.\cite{Serda2016a}

To address these fundamental constraints on widespread adoption, we explored ultra-high dose rate electron-beam irradiation for generating inherently biologically sterile samples with radicals \emph{in situ}: a sample could be prepared in a central (worldwide) facility, rapidly sterilised, and shipped at room temperature around the world prior to use with a polariser in a hospital site. Furthermore, the ability to use powders or crystals of neutral salts, such as alanine, rather than glassing suspensions of trityl radicals and typically organic ketoacids, further dramatically ameliorates the technical requirements of the technique. This approach also opens new classes of molecules to dDNP, as radical formation occurs statistically throughout the sample via well-understood particle-physics interactions, rather than requiring glassed mixtures of disparate radicals (themselves being limited by exclusion from any crystals formed). This allows for effective polarisation of high concentrations of virtually any anhydrous solid-state molecule that would otherwise be challenging to study. All that would be required for dissolution is a neutral sterile dissolution buffer, and, unlike UV-irradiation generated radicals in acids, there is no need for precise control of temperature and magnetic field over many orders of magnitude.\cite{Capozzi2025} Indeed, in this work, we prepared samples for DNP in one country (the UK) and studied them via dissolution-DNP in another (Denmark). 

As anticipated from extensive studies on NA alanine as a dosimeter, e-beam irradiation of NA alanine generated radicals in a dose-dependent manner\cite{Bradshaw1962}, and a similar trend was observed from the \textsuperscript{13}C-enriched samples, saturating at around \SI{70}{kGy}, well above the dose requirement of \SI{25}{kGy} for medical sterilisation under ISO 11137-2.\cite{DIN-Normenausschuss2023, Hoxey} What is surprising, however, is that upon mixing with glycerol the system appears to become partially ordered, and the efficiency of DNP increases substantially. This also generates a clinically useful probe as alanine and glycerol are not miscible in the volume ratios considered here and stable alanine radicals remain present for at least 16 weeks, stored anhydrously in an inexpensive plastic container with silica desiccant gel at room temperature. Furthermore, $g$ changes at low temperature with this system, indicating a fundamental change in the radical electronic environment. We observed no unexpected radiation chemistry byproducts (likely because they spontaneously recombine upon the addition of hot water in the dissolution process, wherein the mixture is melted), and both glycerol and alanine are endogenous biomolecules -- which are safe \emph{in vivo}. This permitted the quantification in the rat kidney of downstream alanine metabolites, directly analogously to the reported biomedical uses of [1-\textsuperscript{13}C]alanine hyperpolarised with trityl radicals.\cite{Park2017,Hu2011,Nielsen2020,Radaelli2020a, Viswanath2021a}

This polarisation enhancement is difficult to quantitatively understand. We have discovered a unique DNP mechanism that is not explained by conventional models yet produces clinically relevant levels of polarisation in an extensively studied biomolecule. The alteration of the frequency sweep profile of nuclear enhancement with radiation dose we hypothesise may be due to some degree of cooperativity between electrons. The temperature-dependent electronic $g$-shift we observed in the EPR data is consistent with magnetic ordering in our system. Indeed, spontaneous spatial ordering in neutral alanine crystallites has previously been reported to occur in strong magnetic fields,\cite{Matsumoto2011a} arising out of magnetic anisotropy of the L-alanine unit cell and its comparatively large (in magnitude) $\chi_m$;\cite{Wang1996, Wang2002, WANG2010} and zirconium-nitrate doped $L$-alanine single crystals form magnetic order and have been used as nonlinear optical devices.\cite{Suresh2020}

While traditional paramagnetic systems maintain independent spin behaviour even at low temperatures, our partially ordered alanine crystallites dispersed in glycerol could exhibit more complex magnetic phenomena as radical centres become close enough to develop short-range magnetic correlations at low temperatures and high radiation doses within individual crystallites. %
 The glycerol matrix in this scenario could function analogously to domain boundaries in ferromagnetic materials. By physically separating the alanine crystallites, glycerol creates discontinuities in the exchange pathways, limiting the correlation length of any magnetic ordering, but permitting thermodynamically driven spin diffusion through weak coupling provided by proton/carbon spins in the matrix. This is conceptually similar to the role of Bloch walls in ferromagnetic materials, though operating at the nano- or microscale between distinct crystallite domains rather than within a continuous bulk material. This model could potentially explain several of our observations:

Firstly, the spatial confinement of magnetic correlations to individual crystallites would naturally lead to anisotropic magnetic responses dependent on crystallite orientation, consistent with our texture analysis showing preferential alignment. Each crystallite may develop its own local magnetic environment influenced by its specific orientation relative to the external field.

Secondly, this partial magnetic ordering could modify the effective field experienced by nuclear spins, altering the resonance conditions for polarisation transfer for that microscopic environment. In conventional DNP models, polarisation transfer is mediated by electron-nuclear dipolar interactions under well-defined frequency matching conditions (e.g.~$\omega_e \pm \omega_N$ for the solid effect) and the angular orientation terms in the dipolar interaction Hamiltonian ($3\cos^2\theta-1$) are assumed to be spatially averaged in a glass. This regionality in orientation together with a transition from individual to collective behaviour would explain why the best conventional theoretical models of DNP fail to predict our observations at higher radical concentrations, as they assume only electron-nuclear interactions and/or are based on thermodynamic arguments assuming some degree of spatial homogeneity and good thermal contact. They do not account for many-body effects that emerge when multiple electrons (arising from multiple radicals in close proximity) interact collectively.

Thirdly, the glycerol ``boundaries'' could create a heterogeneous distribution of magnetic environments throughout the sample. Radicals near the glycerol interface might experience different effective fields than those in crystallite interiors, leading to a distribution of resonance conditions that broadens and shifts the optimal DNP frequency. This domain-like structure might also explain the effectiveness of frequency-swept microwave pulses in our system. By sweeping across a range of frequencies, we could be sequentially addressing different sub-populations of radicals with distinct magnetic environments, each contributing to the overall polarisation enhancement. The constructive summation of these contributions could explain the superior performance compared to fixed-frequency irradiation, even though the frequency swept range is small compared to the overall EPR lineshape. It also provides a mechanism for the biexponential growth of polarisation, which could be ascribed as being that within each crystallite occurring quickly, and that between them occurring through the dilute matrix of natural abundance carbon-13 spins in glycerol. These results are overall surprising, and future work will help elucidate a mechanism behind them.  They represent a new paradigm in DNP, where a partial spatial ordering of spin systems creates emergent behaviour not predicted by isolated electron models of molecular physics.

The levels of nuclear enhancement we achieved are several orders of magnitude larger than those reported for ionising-radiation radical induced DNP previously by $\gamma$-rays by Giannoulis,\cite{Giannoulis2024} and are comparable with the optimal trityl, narrow line-width radicals used clinically.  For a non-irradiated alanine sample, the polarisation level at the point of dissolution was 19.2\% and the nuclear T\textsubscript{1} was \SI{40.8}{s}. For reference, a previous study where alanine was also hyperpolarised with the trityl radical yielded a liquid state polarisation of 12.6\% at the point of dissolution and an \textit{in vitro} T\textsubscript{1} of \SI{41.5}{s} at \SI{3}{T}\cite{Hu2011}. Our irradiated samples produced comparable results; polarisation levels of 17.7\% at the point of dissolution were achieved, while the extracted nuclear T\textsubscript{1} relaxation times were \SI{77.3}{s} at \SI{1.4}{T}. The nuclear T\textsubscript{1} relaxation time was prolonged due to the fact that the irradiation-generated radicals quench upon dissolution. Future work could explore the direct deuteration of the \textsuperscript{13}C-enriched alanine to potentially enhance polarisation levels as previously observed for other molecules hyperpolarised by dDNP\cite{Rooney2023}. Crucially, the pH of the irradiated alanine samples (approximately 7) also made them suitable for \textit{in vivo} trial with dissolution in water; we note that neutral salts are readily crystallised and substances such as sodium pyruvate likely will be readily amenable to this technique

Beyond the physical novelty of our approach, our \emph{in vivo} experiments demonstrate significant translational potential. Electron-irradiated \textsuperscript{13}C-alanine not only matched but in some aspects exceeded the performance of conventional preparations using exogenous radicals. It is worth considering how \textsuperscript{13}C-alanine could compare to \textsuperscript{13}C-pyruvate, the most commonly used metabolic probe with dDNP and undergoing clinical trials. Whilst pyruvate polarises very well with dDNP and is rapidly metabolised to lactate, alanine, and bicarbonate,  it has been noted that the pyruvate/lactate ratio is not reflective of intracellular redox state following HP pyruvate infusion\cite{Zanella2021}. Upon switching the injectable substrate from \textsuperscript{13}C-pyruvate to \textsuperscript{13}C-alanine, Hu et al. recorded more than a ten-fold increase in the \textsuperscript{13}C-lactate/\textsuperscript{13}C-pyruvate ratio\cite{Hu2011}. The difference was suggested to be due to the uptake of \textsuperscript{13}C-pyruvate being limited by perfusion and downstream flux through monocarboxylate transporters during the acquisition window. An incomplete uptake of \textsuperscript{13}C-pyruvate would mean contributions from an extracellular \textsuperscript{13}C-pyruvate pool in the measured \textsuperscript{13}C-pyruvate signal. Furthermore, any extracellular \textsuperscript{13}C-pyruvate present in plasma can be converted into \textsuperscript{13}C-lactate by red blood cells that contain lactate dehydrogenase, leading to an ``extracellular'' \textsuperscript{13}C-lactate pool\cite{Park2017}. Signals arising from intracellular and extracellular pools cannot be easily distinguished in hyperpolarised MR, leading to the proposed use of \textsuperscript{13}C-alanine for determining the \textsuperscript{13}C-lactate/\textsuperscript{13}C-pyruvate ratio. For \textsuperscript{13}C-alanine to be converted into \textsuperscript{13}C-pyruvate, and subsequently \textsuperscript{13}C-lactate, it must first be exposed to the intracellular enzyme alanine transaminase. Thus, the \textsuperscript{13}C-pyruvate signal detected following the injection of \textsuperscript{13}C-alanine would be intracellular in origin. Moreover, after the metabolism of alanine is observed and quantified, the cellular redox potential can be easily determined \cite{Park2017} from the known relation  $K_\text{eq}=\frac{[\text{Pyruvate}]\,[\text{NADH}] [\text{H}^+]}{[\text{Lactate} ][\text{NAD}^+]}$ where $K_\text{eq}=1.11\times10^{-11}$ M.

Our electron irradiation approach uniquely combines this biochemical advantage with the practical benefits of \emph{in situ} radical generation and spontaneous quenching, providing a powerful new tool for quantifying intracellular redox potential. It is also the first to explicitly consider partially ordered solids as substrates for biomedical dDNP rather than amorphous glasses that are commonly used with the technique (and, indeed, often listed as a requirement for its success; \cite{LillyThankamony2017a}) While this technique could be extended to other substrates, \textsuperscript{13}C-alanine represents an ideal initial candidate due to its metabolic significance, detailed quantum-mechanical study, and demonstrated biochemical and clinical relevance. This approach may reduce the technological, regulatory, and economic barriers that have constrained hyperpolarised MRI primarily to specialised research centres, potentially expanding access to this powerful diagnostic capability.

\subsection{Limitations}

We should note that several limitations should be acknowledged in this work: notably, our mechanistic understanding of the DNP process in this system remains incomplete and in need of extension via theoretical modelling. The unusual frequency sweep profiles of this highly efficient polarisation process hint at cooperativity and physics beyond conventional DNP models, but a comprehensive theoretical framework that quantitatively predicts these behaviors is lacking. Additional spectroscopic studies with variable temperature, field strength, and radical concentration would help elucidate the underlying mechanisms. 
Additionally, while we demonstrated stable radical formation in alanine, the applicability of this approach to other metabolically relevant molecules requires further investigation. Different molecular structures may respond differently to electron irradiation, potentially generating distinct radical species with varying stability and DNP efficiency. The specific arrangement of alanine in a crystalline lattice likely contributes to the observed radical stability and its partial ordering in glycerol (analogous to a nemantic liquid crystal), and molecules with different crystal structures, or those that do not readily crystallise, may present challenges. Furthermore, the comparatively high radiation doses considered may embrittle plastics used in the construction of fluid paths, but this phenomenon is well-characterised, and many medical plastics are routinely sterilised with comparable doses.\cite{InternationalAtomicEnergyAgency2021,White2013}

Finally, the highest nuclear polarisation levels achieved with our approach ($\sim20\%$) remain lower than those reported for optimised pyruvic acid preparations with trityl radicals ($>50\%$). While sufficient for \emph{in vivo} imaging, higher polarisation would  improve SNR and potentially allow detection of lower-concentration metabolites. Further optimisation of irradiation parameters, alanine crystallite size, and DNP conditions might narrow this gap. Likewise, owing in part to this limitation, our \emph{in vivo} experiments were limited to proof-of-concept demonstrations in a small number of healthy animals. The performance of irradiated alanine in disease models, particularly its sensitivity to pathological alterations in redox potential, requires validation in relevant preclinical models before clinical translation.

\section{Materials and Methods}
\small
\subsection{Sample Preparation}

The samples used in this study can be categorised into three main groups: alanine containing endogenous irradiation-generated radicals, alanine mixed with the exogenous trityl radical OX063 at a concentration of \SI{38}{mM}, and pyruvate with \SI{30}{mM} AH1111501. These exogeneous radicals differ only by an exterior methyl `R'-group in their chemical structure and have almost identical electronic properties; these concentrations of both trityl radicals were chosen to reflect reported optimal conditions for DNP for both molecules. 

To prepare the former, dry [1-\textsuperscript{13}C]L-alanine powder (Isotec and Cambridge Isotope Laboratories, Inc) was irradiated using an ultra-high dose rate 6 MeV electron linear accelerator operating in pulsed mode, delivering \SI{25}{Gy} of dose to the sample in each of the \SI{4}{\micro s} pulses, with a pulse repetition rate of \SI{25}{Hz} and an approximate beam current of \SI{400}{mA}. The dose of irradiation was varied between 10 -- \SI{100}{kGy} depending on the desired radical concentration and was linear with the total number of pulses delivered. The dosimetry and beam delivery was controlled using a beam monitor, which has been described in detail elsewhere,\cite{Berne2021a, Vojnovic2023} and calibrated against radiochromic film (EBT-XD, Ashland Inc., Covington, KY, USA). The total time for irradiation was deliberately lengthened to avoid linac heating; approximately \SI{2}{g} was irradiated in \SI{160}{s}. The irradiated powder was then used as the main component for two different sample types: either an irradiated dry powder or an irradiated powder/anhydrous glycerol mix with a 1:1 w/w ratio.

The alanine preparation containing OX063 trityl radical (molecular weight = \SI{1427}{g/mol}; GE Healthcare) comprised \SI{1.276}{g} of [1-\textsuperscript{13}C]L-alanine mixed in 8 ml of water and \SI{2}{ml} of \SI{11.65}{M} HCl. This mixture was freeze-dried, and approximately \SI{1.5}{g} of alanine hydrochloride yielded. The alanine hydrochloride was then dissolved in \SI{2.25}{ml} DMSO with the assistance of heating. \SI{180}{mg} OX063 trityl was added to achieve approximately \SI{38}{mM} radical concentration. \SI{50}{\micro l} of this mixture was pipetted into a sample cup.

For the pyruvate sample, \SI{14}{M} [1-\textsuperscript{13}C]pyruvic acid was mixed with the appropriate mass of AH111501 trityl (GE Healthcare; CID 11607875) to achieve a \SI{30}{mM} radical concentration, of which \SI{18}{\micro l} was pipetted into a sample cup.

\subsection{Dynamic Nuclear Polarisation}

Samples were polarised at 3.35~T or 6.7~T using a HyperSense (Oxford Instruments) or SpinAligner (Polarize), respectively. Further details are provided in the \SIRef{si:DNP}.

\subsection{In Vitro Experiments}

The polarisation levels and nuclear T\textsubscript{1} relaxation times of samples were recorded using a benchtop NMR spectrometer at 1.4 T (SpinSolve 60\textsuperscript{ULTRA} Carbon, Magritek). The accompanying software also enabled the level of polarisation to be estimated at the point of dissolution.

\subsection{In Vivo Experiments}

Two healthy male Sprague-Dawley rats were injected with one shot of each type of hyperpolarised alanine sample via a tail vein catheter. The time between each injection was approximately two hours. The respiratory rate and body temperature of each rat were monitored using an MRI compatible small-animal monitoring system (Small Animal Instruments Inc, USA) whilst the rats were under anaesthesia with 2.5-3\% sevoflurane in \SI{1}{l/min} medical air. Both rats were scanned using a dual-tuned \textsuperscript{13}C/\textsuperscript{1}H volume rat coil at \SI{3}{T} (GE MR750, GE Healthcare). One rat was scanned using a slice-selective spectroscopic acquisition, the other imaged using a 2D MR Spectroscopic Imaging (MRSI) sequence, EPSI, using fidall. 

Prior to these scans, a urea phantom was used to calibrate the applied flip angles, and proton images were acquired for anatomical reference. For the slice-selective spectroscopic acquisition, \SI{80}{mm} slice thickness was used to ensure the whole abdomen of the rat was imaged. \SI{1}{ml} of hyperpolarised sample was injected over approximately 5 s, and the scan started upon the start of the injection. A \SI{30}{\degree} flip angle was applied with a repetition time (TR) of \SI{3}{s} for a total of 5 minutes (100 time points). For the MRSI, \SI{1}{ml} of hyperpolarised substrate was injected over approximately \SI{5}{s}, and the scan started upon the start of the injection. A $16\times16$ matrix was chosen, and a \SI{15}{\degree} FA was applied with a TR of \SI{64}{ms} for a total of 2 minutes (8 time points and approximately \SI{16}{s} per time point). The pH of the injected solutions was measured to ensure physiological compatibility, with values ranging from 6.7 to 7.4, well within the acceptable range for intravenous administration. All animal experiments were conducted following appropriate independent ethical review and under appropriate licensing regimes in Denmark.

\subsection{Electron Paramagnetic Resonance}

An X-band continuous wave (CW) spectrometer (EMXmicro, Bruker BioSpin; Super High Sensitivty rectangular cavity probe head) was used to record EPR spectra at room temperature. For recording spectra at \SI{5}{K} and an estimate of $T_{1e}$, an X-band CW spectrometer (E680, Bruker BioSpin) was used. Magnesium Oxide powder, a common standard in EPR,\cite{Burghaus1992} was used as a frequency reference.

Samples that comprised dry powder were directly loaded into \SI{4}{mm} thin wall quartz EPR tubes, and their mass recorded. For repeated measurements, samples were rotated about their own axis by approximately 25--\SI{35}{\degree}. Liquid state samples were first loaded in \SI{2}{mm} diameter capillary tubes, which in turn were loaded into the EPR tubes.

Spin counts were determined from EPR spectra using an in-house developed MATLAB script. The script incorporated baseline correction, normalised for differences in resonator $Q$-factor, and integrated the recorded spectra.

\textsuperscript{13}C spectra were fitted via a constrained Bayesian optimisation routine\cite{Martinez-Cantin2014}. The line widths and shapes of spectra were also compared to models built with the `pepper' function in EasySpin, a validated QM spectral simulation library designed for EPR.\cite{Stoll2006}

\subsection{X-ray Diffraction}

XRD was conducted using copper $K_\alpha$ X-rays, of wavelength \SI{1.5405982}{\angstrom} and \SI{1.544497}{\angstrom}, in an Empyrean (Malvern Panalytical) diffractometer. XRD samples comprised dry alanine powder or an alanine powder/glycerol mix (1:1 w/w). The Cu $K_\beta$ wavelength was removed from the diffraction signal with a Ni filter placed in front of the PIXcel1D detector. A set of programmable divergence slits were used for the incident optic, set to illuminate a fixed surface area of \SI{1}{cm}$^2$ across the angular range of the scan. During the scans, the samples were rotated about the azimuthal axis at a frequency of \SI{0.5}{Hz} to fully characterise the crystallite distribution.

XRD data were compared to a previously reported crystal structure of alanine, obtained via neutron scattering\cite{Wilson2005}, and a Rietveld refinement was undertaken in Maud using a spherical harmonic texture basis\cite{Lutterotti1999}. Further details are provided in the \SIRef{SI:textureAnalysis}.

\emph{[6167 Words]}


\section{Figures and Tables}

%
%
%
%
%


\clearpage 

%
\bibliography{Bibliography} 
\bibliographystyle{sciencemag}
\clearpage 

%
%
%
%
%
%


\section*{Acknowledgments}
\paragraph*{Funding:}
We would like to acknowledge support from the Novo Nordisk Foundation (NNF21OC0068683); the British Heart Foundation (refs. RG/11/9/28921, RE/13/1/30181, FS/19/18/34252, and FS/14/17/30634); the Wellcome Trust (221805/Z/20/Z); the Engineering and Physical Sciences Research Council (EPSRC) and Medical Research Council (MRC) (EP/L016052/1); and the Colleges of St. Hugh's and Somerville in the University of Oxford. We would furthermore like to acknowledge the University of Oxford British Heart Foundation Centre for Research Excellence (RE/13/1/30181) and the NHS National Institute for Health Research Oxford Biomedical Research Centre programme. 

The views expressed are those of the authors and not necessarily those of the NIHR or the Department of Health and Social Care. We also gratefully acknowledge Cambridge Isotope Laboratories, Inc, who provided a gift of [1-\textsuperscript{13}C]alanine to J.J.M. and Z.R. We also thank Christiane Timmel for useful discussions about quantitative EPR spectroscopy. JJM would additionally like to thank Professor Elspeth Garman and Professor Ian Carmichael for both dinner and many useful discussions about radiation chemistry in crystals.  The practical assistance of the Department of Physics Mechanical Workshops is also kindly noted. 
\paragraph*{Author contributions:}
C.H.E.R. and J.Y.C.L. conceptualised research, conducted experiments, analysed data, and drafted the manuscript. E.S.S.H., N.V.C., and D.A.D. assisted with sample preparation and \emph{in vivo}/high field DNP experiments; B.W.C and S.S. assisted with low-field DNP experiments. K.P., I.T., B.V., C.H.E.R. undertook electron linear accelerator experiments for sample irradiation, including design and manufacture of equipment. J.L, C.H.E.R. and J.J.M undertook X-ray diffraction experiments and their interpretation. W.M., and A.B. provided expertise in EPR spectroscopy and assisted with data interpretation. Z.R. assisted with modelling. L.B.B. and C.L. supervised the \textit{in vivo} experiments and contributed to experimental design. J.Y.C.L, D.J.T. and J.J.M. conceived and supervised the project, secured funding, analysed data, and finalized the manuscript. All authors reviewed, edited and approved the final manuscript.
\paragraph*{Competing interests:}
JYCL, CL, JJM, CHER and DT are listed as authors on a filed patent (EP24168322.6) pertaining to the use of ultra-high dose rate electron irradiation for preparing hyperpolarised contrast agents. All other authors have no conflicts of interest to declare. 
\paragraph*{Data and materials availability:}
Crystallographic data pertaining to this project are enclosed as CIF files as SI to this paper. Materials (i.e.~ultra-high dose rate irradiated labelled metabolites) are available to others upon the signing of a collaboration and material transfer agreement. All other data are available in the main text or the supplementary materials.

\subsection*{Supplementary materials}
Materials and Methods\\
Supplementary Text\\
Figs. S1 to S9\\
Table S1\\
References \textit{(7-\arabic{enumiv})}\\ 
Data S1


\newpage


\renewcommand{\thefigure}{S\arabic{figure}}
\renewcommand{\thetable}{S\arabic{table}}
\renewcommand{\theequation}{S\arabic{equation}}
\renewcommand{\thepage}{S\arabic{page}}
\setcounter{figure}{0}
\setcounter{table}{0}
\setcounter{equation}{0}
\setcounter{page}{1} 


\begin{center}
\section*{Supplementary Materials for\\ \scititle}

    Catriona~H.~E.~Rooney \emph{et al}.
    \small$^\ast$Corresponding author. Email: jack.miller@physics.org\and
\end{center}

\subsubsection*{This PDF file includes:}
Materials and Methods\\
Supplementary Text\\
Figures S1 to S9\\
Table S1 \\

\subsubsection*{Other Supplementary Materials for this manuscript:}
Data S1\\

\newpage


\section{Supplemental Experimental Methods}
\subsection{Irradiation}
\label{SI:irradiation}
Full technical details of the linac are published elsewhere and much of its construction details are freely available at \url{https://users.ox.ac.uk/~atdgroup/technicalnotes/}. 

\begin{figure}
    \centering
    \includegraphics[width=0.7\linewidth]{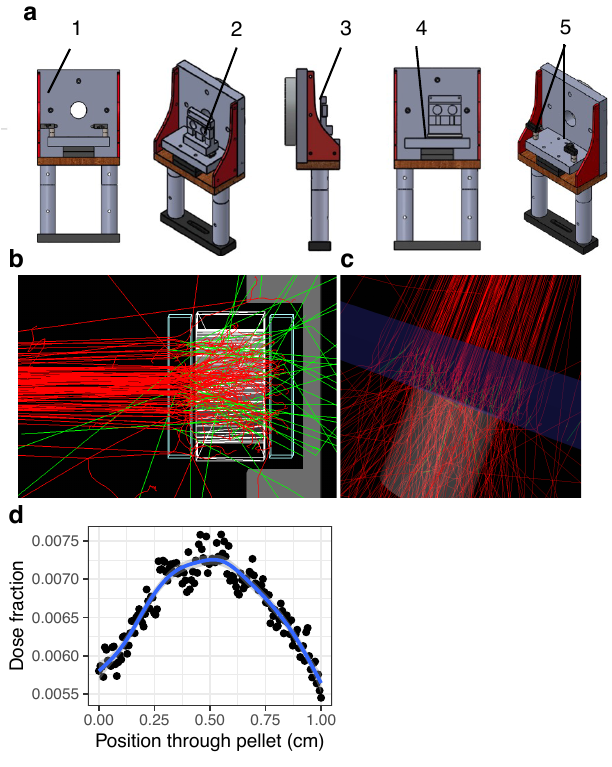}
    \caption{\textbf{a:} A sample holder for the irradiation of alanine samples using a 6 MeV high-dose rate electron linear accelerator was designed. \textbf{a} Renderings of the sample holder comprising (1) an aluminium endplate with a 5 cm diameter aperture for the electron beam, (2) a rectangular block of aluminium --- with two cylindrical holes each acting as a sample space --- interposed between two glass microscope slides and secured upright with microscope slide spring clips to (3) an L-shaped aluminium block attached onto (4) a high precision linear slider. (5) Two stoppers were positioned either side of the slider to facilitate the positioning of each sample. \textbf{b, c}: Simulated Monte-Carlo radiation transport characteristics of the designed target demonstrating appropriate dose deposition inside the compressed alanine pellet (red denotes leptons; green photons); together with \textbf{(d)} predicted dose deposition as a function of depth, normalised such that the total dose is unity.}
    \label{fig:SIsample_holder}
\end{figure}

Renderings of the sample holder used to hold the sample during the irradiation are shown in panel (a) of Fig.~\ref{fig:SIsample_holder}. To produce the samples, a block of aluminium with two sample spaces was loaded with dry alanine powder. The powder was secured in place by taping two ordinary glass microscope slides either side of the aluminium block. The glass-aluminium-glass sandwich was held in an upright position by microscope slide spring clips within another aluminium component that itself was connected to a high precision linear slider. The integration of the slider and two stoppers into the design facilitated the centring of each sample space within the beam path. To increase the homogeneity of the irradiation exposure across the total volume of each sample space, the glass-aluminium-glass sandwich was flipped 180\textdegree{} after half of the total desired irradiation dose had been delivered. Irradiation doses over the range 10--100 kGy were applied to produce samples with a range of different endogenous radical concentrations. 

A radiation transport simulation was undertaken in TOPAS,\cite{Faddegon2020, Perl2012} an open-source radiation transport code optimised for medical physics applications based on Geant4, written by CERN.\cite{Agostinelli2003} As shown in Fig.~\ref{fig:SIsample_holder}\textbf{b, c}, this was used  to optimise dose distribution across the sample with regards to the `pile up' dose deposited either in the glass coverslips or aluminium target; the preferred deposition function is that of a `top hat' within the sample but it was found that appropriate spacing and modulation of design produced a reasonable distribution function (Fig.~\ref{fig:SIsample_holder}\textbf{d}). In order to obtain a more homogeneous dose distribution within the sample, we rotated the sample by \SI{180}{\degree} in the middle of the irradiation process; after irradiation, the sample was mixed by hand, further homogenising it. 

\subsection{DNP}
\label{si:DNP}

DNP was performed either using an Oxford Instruments Hypersense polariser at \SI{3.35}{T} or a customised SpinAligner \SI{6.7}{T} polariser. In both cases either microwave sweep profiles were obtained by iterating over microwave irradiation frequency, or polarisation build-up curves in which a low flip angle pulse was repeatedly played and polarisation monitored. 

Reported time-constants for the DNP build-up profiles were fitted to a biexponential curve given by the equation

\begin{equation}
y = P(A(1-e^{-\frac{\text{time}}{T}})+(1-A)(1-e^{-\frac{\text{time}}{t}}))
\end{equation}

where $P$ is the build-up constant, $A$ and $1-A$ are the relative weightings of the fast and slow components of the polarization build-up, and $T$ and $t$ are the fast and slow components of the build-up times. 

For the Hypersense, microwave frequency sweeps were acquired for each sample between 93.75 to \SI{94.195}{GHz} in \SI{5}{MHz} steps with a two minute build-up per frequency step. The microwave power was held constant at \SI{100}{mW} for all microwave frequency sweeps and build-ups. The build-up of polarisation was subsequently monitored at the peak frequency of the microwave frequency sweep every three minutes using a low flip-angle readout.

Based on previous optimization work and the use of a narrow line width radical, the SpinAligner was used without microwave modulation for the non-irradiated samples. For the irradiated samples, microwave modulation was applied over a range of 25 MHz about the given centre frequency, sinusoidally varied with the frequency of microwave irradiation occurring at a rate of \SI{1}{kHz}. This applies a weak modulation and attempts to invert a greater proportion of electron spins; it also is recommended by the microwave source manufacturer in order to promote greater thermal stability of their system.

For the Spinaligner, microwave frequency sweep profiles were acquired using 32 averages, an RF transmit pulse width of \SI{2}{\micro s}, and a TR of \SI{30}{s} for irradiated samples or a TR of \SI{10}{s} for all other sample types.
The build-up proﬁles were acquired using \SI{2}{\micro s} transmit RF pulses with a TR of \SI{60}{s} at \SI{20}{mW} microwave power and no averaging for all sample types. The transmission power of the RF pulses was \SI{32}{dBm}, although the exact $B_1$ generated inside the sample cavity is poorly characterised. 

The NMR spectra recorded during microwave frequency sweeps and polarization build-ups were quantified by peak integration in MATLAB (ver.~2023a; The Mathworks, Inc). Polarization estimates were obtained in the liquid state both hyperpolarised and at thermal equilibrium, and then back-calculated using the measured $T_1$ and known time of dissolution to estimate solid-state polarisation; for samples with low nuclear polarization in the solid state (where relative error post dissolution is likely higher or sufficiently high to be inaccurate) a direct linear scaling of ADC values obtained at constant gain with a known polarisation reference were used. 

\subsection{Polarisation characterisation}
\label{SI:builduptimes}

As summarised in Tab.~\ref{tab:pol_constants}, we found that an approximately inverse relationship between build-up time constants and irradiation dose, i.e. the greater the radical concentration, the longer the slow component of the build-up time and the shorter the fast component. 

This is potentially because of the competing effects of increased polarization transfer efficiency and enhanced nuclear relaxation at higher radical concentrations. The shortened fast component likely reflects more efficient initial polarization transfer due to increased electron-nuclear spin coupling, while the extended slow component may indicate limitations in long-range spin diffusion as paramagnetic centres become more densely packed. At higher radical concentrations, the shortened inter-radical distances could impede the cooperative network of nuclear spins necessary for efficient polarization distribution throughout the sample, creating isolated `islands' of polarization around each radical centre. This spatial heterogeneity in the polarization landscape would manifest as different characteristic time constants in the bi-exponential build-up curve. 

At very high radical concentrations, it may be the case that the electronic environment in the crystal is similar to that of a doped semiconductor with a complex band-structure, at which point existing theories of DNP are likely not complete. 

\begin{table}[htbp]
\centering
\caption{Where $T$ is the slow component of the build-up time and $t$ is the fast component of the build-up time.}
\label{tab:pol_constants}
\begin{tabular}{cccc}
    \hline
Radiation dose (kGy) & Build-up Amp: P (a.u.) & $T$ (s) & $t$ (hours) \\
\hline
30 & 1520000 & 22.9 & 0.92 \\
50 & 2280000 & 22.2 & 2.69 \\
70 & 5110000 & 14.9 & 6.88 \\
\hline
\end{tabular}
\end{table}

\subsection{Spin count procedure and supplementary data}
\label{SI:SpinCount}
Quantitative EPR spectra were recorded on an X-band CW EPR spectrometer (EMXmicro, Bruker Biospin GMBH) at room temperature; electronic $T_2$ measurements were performed at \SI{5}{K} on an X-band E680 spectrometer. Magnesium Oxide (MgO) powder, a common standard reference in EPR, was used to account for changes in the magnetic field between experiments due to differences in sample positioning and resonator $Q$ values; it is known that the signal amplitude is proportional to $\chi_m Q \sqrt{\text{Microwave power}}$ in EPR and these parameters were controlled appropriately.\cite{Burghaus1992}
Samples that comprised dry powder were directly loaded into 4 mm thin wall quartz EPR tubes and their mass recorded. For repeated measurements, samples were rotated about their own axis by approximately 25 to \SI{35}{\degree} to ensure adequate powder averaging. Liquid state samples were first loaded in 2 mm diameter capillary tubes which in turn were loaded into the EPR tubes. 

For longitudinal studies observing the stability of radical species, EPR spin counts were recorded on the same day as their irradiation and then two weeks, four weeks, and 16 weeks post irradiation. Dry powder samples were prepared directly in the EPR tubes and rotated about their own axis as above; the mass of empty and filled tubes were recorded and monitored over time and data are corrected for the mass of samples present. Solutions containing glycerol were prepared as stock solutions which were used to fill haematocrit tubes which themselves were inserted into the EPR tubes. Best practices for undertaking quantitative EPR spectroscopy were followed, such as the use of an appropriate resonator, and reproducible packing metrics with impurity controlled materials.\cite{Eaton2010, Mazur2006} EPR data was analysed in Matlab. Spin counts were determined by double numerical integration of the EPR spectra. First, the raw spectra were baseline-corrected using a first-order correction, followed by integration to obtain absorption spectra. Q-value normalization was performed to that using a reference Q-value of 8000 (obtained with the MgO sample) to account for variations in cavity loading. A polynomial fit (order 15) was then applied to specific field regions (317.5--333 mT and 337--350 mT) for secondary baseline correction. The final double integration value was calculated from the fully corrected absorption spectrum, with verification performed by comparing the numerical derivative of the absorption spectrum against the original intensity data. Full width at half maximum (FWHM) values were also calculated for each spectrum to characterize line broadening effects related to radical concentration and interactions. 

For conversion between spin counts and radical concentrations, a calibration curve was established using TEMPO samples of known concentration, allowing quantitative determination of radical concentration in alanine samples.

For use with quantitative modelling studies, electronic $T_1$ and $T_m$s were collected via use of an inversion recovery pulse sequence ($T_1$) and $T_m$ (approximately $T_2$) by a 2-Pulse Echo Decay method. An approximate estimate of the electronic $T_1$ of the SAR at 5 K in glycerol was \SI{12.4}{s} and $T_m\approx\SI{371}{ns}$. For comparison, a ``literature prep'' (non-irradiated [1-13C]L-alanine, 18.94M NaOH, DMSO, 15mM OX063 and 0.3mM Dotarem gadolinium) had $T_{1e} = \SI{41}{ms}$, $Tm = \SI{222}{ns}$, ``literature prep without gadolinium''  had $T_{1e} = \SI{4.6}{s}$ and $T_m = \SI{181.6}{ns}$.

\begin{figure}[htbp]
\centering
\includegraphics[width=0.8\textwidth]{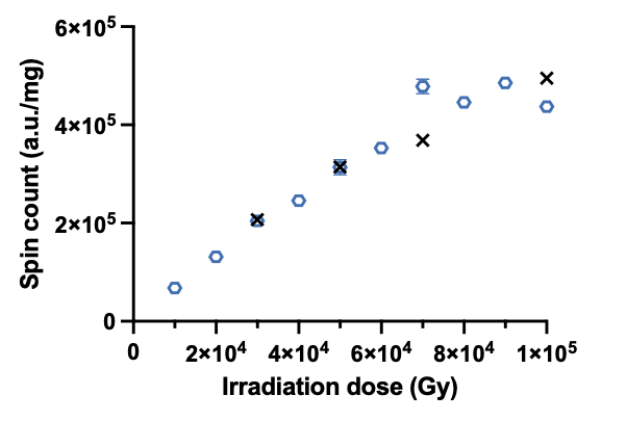}
\caption{Spin counts recorded from \textsuperscript{13}C-alanine (black crosses, n=3) were consistent with those recorded from NA alanine (blue hexagons, n=4) at irradiation doses 30, 50, 70, and 100 kGy.}
\label{fig:supp1}
\end{figure}

As shown in Fig.~\ref{fig:supp1}, no significant difference in dose-response curves was observed with [1-\textsuperscript{13}C]alanine compared to natural abundance alanine. This is largely expected because the predominant method of energy loss at 6 MeV is via Compton scattering, with a cross section on the order of millibarns. In contrast, electron-nuclear interactions such as electro-disintegration or nuclear excitation have microbarn to nanobarn cross sections at these energies, and have been extensively studied\cite{Barreau1983, Cheon1983, Deady1986} as a probe of differences in the radii of \textsuperscript{12}C and \textsuperscript{13}C nuclei (a difference of approximately \SI{0.023$\pm$0.01}{fm}).\cite{Heisenberg1970} Furthermore, the mass-energy absorption coefficients for \textsuperscript{12}C and \textsuperscript{13}C differ by only $\sim$0.08\% due to the marginal difference in electron density per unit mass.\cite{148751} Therefore, any isotopic effect on dose deposition would likely be within experimental uncertainty of our measurements.

\subsection{Alanine SAR EPR Fitting}
\label{SI:AlanineSAROptimisation}

To try to fit the EPR data obtained of irradiated powder 13C alanine, a detailed simulation was undertaken in EasySpin using appropriate previously published hyperfine constants and $g$-tensors for the radical R1 only, which is responsible for the majority of the signal recorded via EPR. The initial fitting model included a spin system with an $S=1/2$ electron and a $g$-tensor consistent with the R1 alanine radical ($g=[2.0041, 2.0034, 2.0024]$). The model incorporated hyperfine couplings for five protons (four $\beta$-protons and one $\alpha$-proton) with starting values derived from NA alanine, plus the $^{13}$C nucleus (summarised below, in Code Listing 1). We employed the \texttt{pepper} function provided by EasySpin\cite{Stoll2006} with a hybrid method for orientational averaging, using an adaptive grid approach with an initial grid size of \texttt{[10 2]} and a minimum grid size of 2. This was progressively refined throughout the optimisation process. Isotropic line broadening from unresolved interactions was modelled with a peak-to-peak linewidth (\texttt{lwpp}) parameter of 0.45~mT. 

The experimental spectra were first-derivative spectra collected at an X-band microwave frequency of approximately 9.4~GHz (with the exact value extracted from experimental parameters) with a field range covering 330-350~mT. Spectra were baseline-corrected prior to fitting by subtracting the mean of the first 128 points.

Fitting was performed using a two-stage optimization approach. First, we employed MATLAB's \texttt{surrogateopt} function with 10,000 maximum function evaluations, exploring a bounded parameter space centred around initial estimates. The objective function combined both spectral and integral matching metrics to ensure proper reproduction of both line shapes and relative intensities. Specifically, we minimized:
\begin{equation}
\min_{\theta} \| \mathcal{I}_{\text{fit}}(\theta) - \mathcal{I}_{\text{exp}} \|_2 + \| \mathcal{S}_{\text{fit}}(\theta) - \mathcal{S}_{\text{exp}} \|_2
\end{equation}
where $\mathcal{I}$ represents cumulative integrals of spectra and $\mathcal{S}$ represents the normalized spectra themselves.

The final fit was further refined using Bayesian optimization with additional iterations (150) to ensure convergence to the global minimum. The fitting achieved satisfactory spectral reproduction with the dominant R1 species, though slight deviations suggest potential minor contributions from R2 and R3 species or other structural heterogeneities in the $^{13}$C-labelled samples.

\subsubsection{Code Listing 1}

\begin{verbatim}
%See doi: 10.1021/jp972158k for details 
published_eigenvals=[2.0041 2.0034 2.0024];
published_eigenvecs=[0.828 0.030 0.559;
    -0.428 0.678 0.598;
    0.362 0.735 -0.574];
g_tensor=published_eigenvecs*diag(published_eigenvals)*inv(published_eigenvecs);

%Anisotropic couplings of the alanine R1 radical
A1 = [-87.8 -52.5 -27.8];          % H-alpha principal hfc values (MHz)
A2 = [74.7 67.6 67.3];             % H-beta principal hfc values (MHz)
A3 = [5.1 -2.1 -2.6];              % H-gamma principal hfc values (MHz)
%(The system is [A1;A2;A2;A2;A3]; for the five different protons)

%Euler angles are
V1 =[-0.6420   1.3484    0.5622];  % H-alpha Euler angles
V2 =[0.0574    2.2199    1.9514];  % H-beta Euler angles
V3 =[2.9103    2.5350   -1.6269];  % H-gamma Euler angles

Sys = struct('S',1/2,'g',published_eigenvals);
Sys = nucspinadd(Sys, '1H', A1,V1); %HAlpha 
Sys = nucspinadd(Sys, '1H', A2,V2); %HBeta
Sys = nucspinadd(Sys, '1H', A2,V2); %HBeta
Sys = nucspinadd(Sys, '1H', A2,V2); %HBeta 
Sys = nucspinadd(Sys, '1H', A3,V3); %HGamma 
%We now add the 13C below and numerically solve for its A and V. 
\end{verbatim}

\subsection{X-ray diffraction texture analysis}
\label{SI:textureAnalysis}

\begin{figure}
    \centering
    \includegraphics[width=\linewidth]{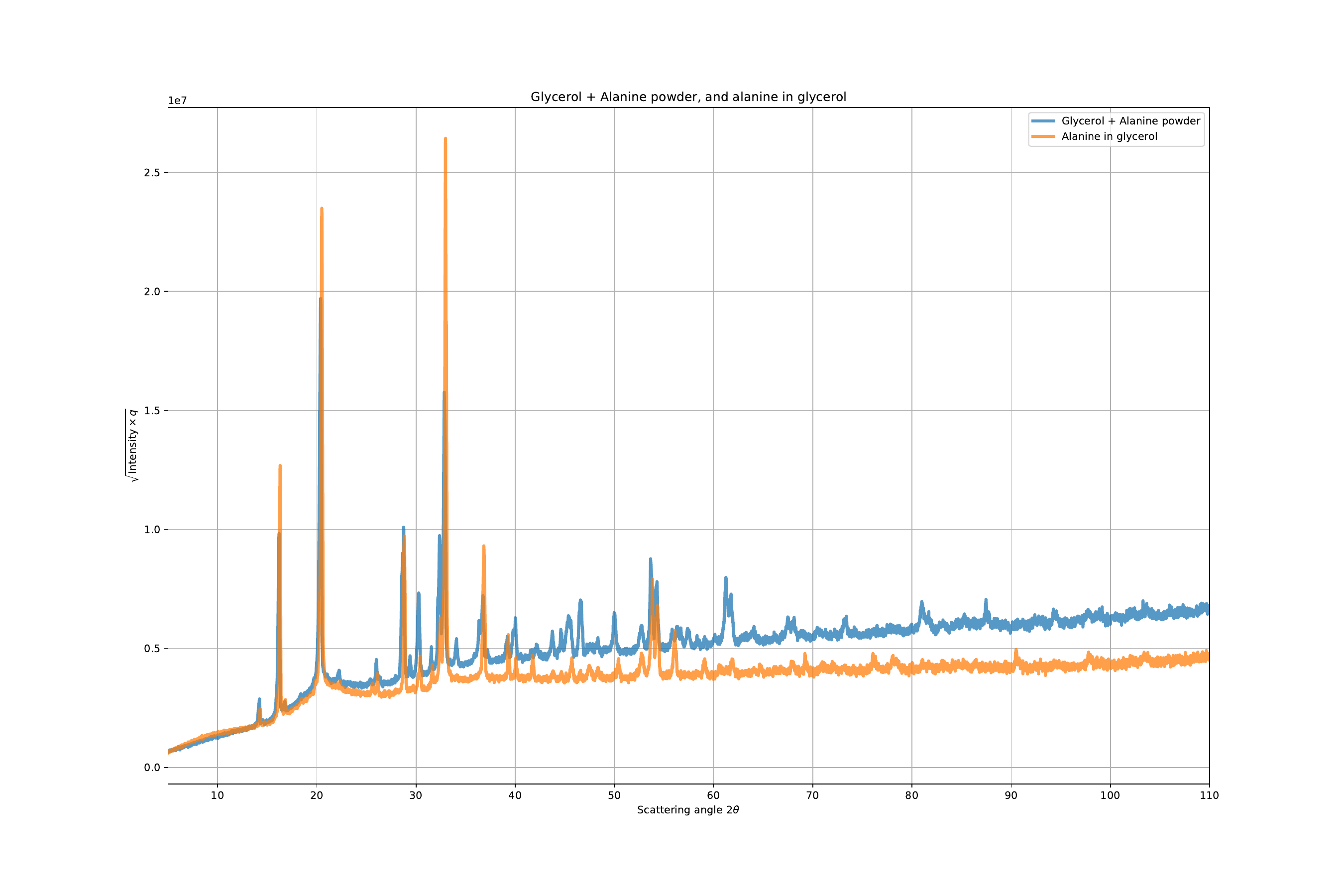}
    \caption{A comparison of (1) the algebraic sum of alanine power and separately acquired glycerol XRD data (blue), and (2) the separately acquired spectrum from an alanine powder + glycerol mixture (orange). A clear change in crystallographic texture is evident from the variation in the relative intensity of the Bragg diffraction peaks. The presence of these peaks indicates that the alanine crystallites have not dissolved in glycerol.}
    \label{fig:powder-comparison}
\end{figure}

High-quality diffraction data was obtained with a signal-to-noise ratio of approximately $10^{12}$ with visually clear differences between the two samples, irrespective of the higher (expected) polynomial background due to excess liquid glycerol in the case of the sample with glycerol (Fig.~\ref{fig:powder-comparison}). 
Quantitative Rietveld refinement was performed using MAUD (Materials Analysis Using Diffraction, version 2.9993), a validated Java-based crystallographic software. Initially, the known orthorhombic crystal structure of L-alanine was employed as a starting model. The following parameters were systematically refined, in order:

\begin{enumerate}
    \item Background (polynomial parameters)
    \item Sample displacement and zero-offset (\SI{-0.081}{\degree} $\pm$ \SI{0.0009}{\degree})
    \item Unit cell parameters
    \item Overall temperature factor (B-iso = $3.00 \pm $\SI{0.15}{\angstrom}$^2$)
    \item Instrumental profile parameters (Caglioti function for peak broadening)
    \item Crystallographic texture
\end{enumerate}

For both samples, the refinement yielded satisfactory crystallographic $R$-factors, with the glycerol-dispersed sample showing $R = 10.2\%$ and weighted $R_w = 14.6\%$, compared to $R = 12.8\%$ and weighted $R = 18.6\%$ for the pure powder. These values indicate good agreement between the experimental data and the structural model after accounting for texture.

\begin{figure}[htbp]
    \centering
    \includegraphics[width=0.7\linewidth]{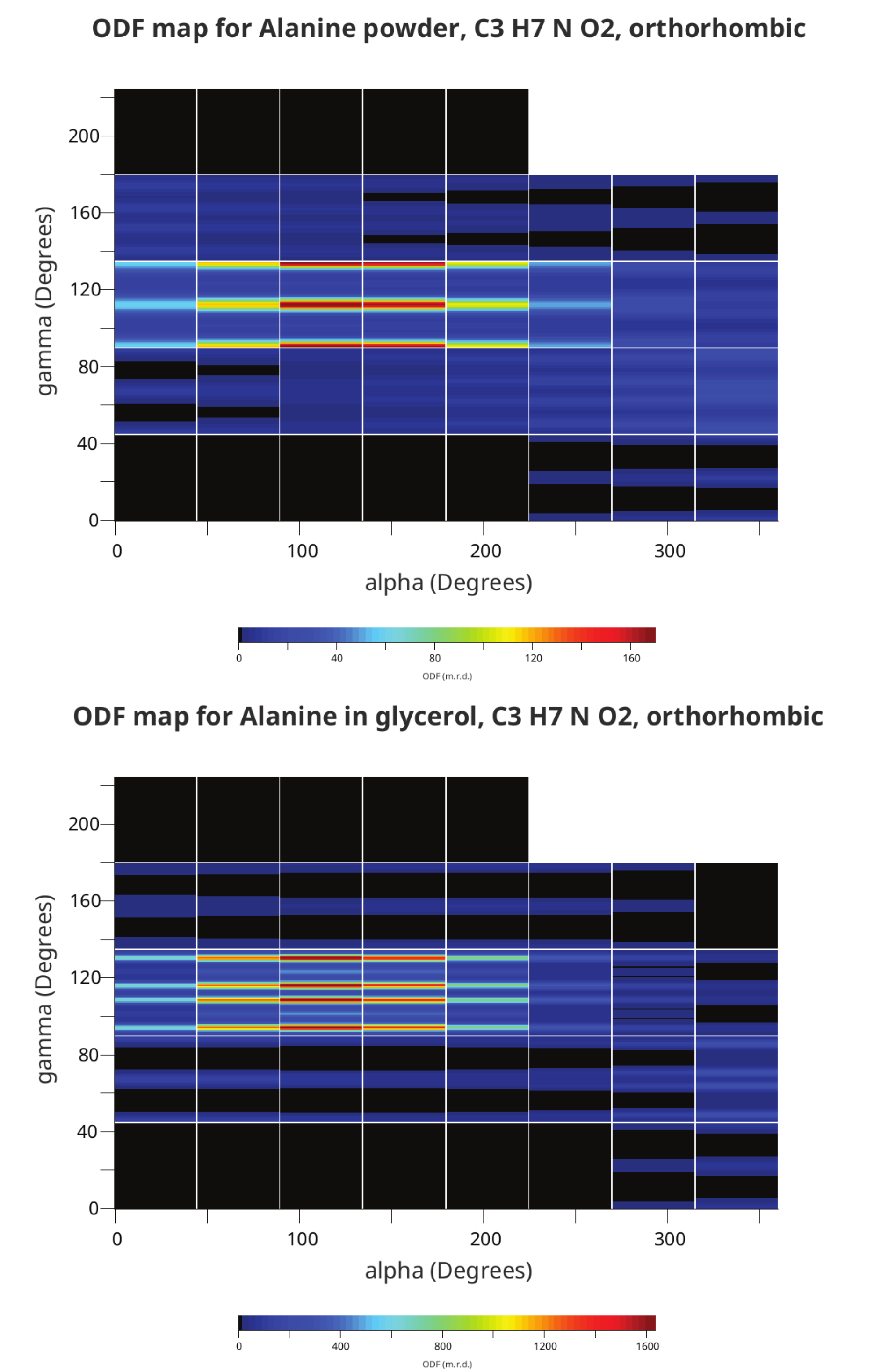}
    \caption{Orientation distribution functions for alanine and alanine/glycerol mixtures show a clear orientation preference that is more than a factor of ten times greater than for alanine alone. }
    \label{fig:ODFs}
\end{figure}

Texture analysis was carried out using the E-WIMV algorithm implemented in MAUD with a spherical harmonic approach. This method involves reconstructing the orientation distribution function (ODF) from the diffraction data to quantify the degree of preferred crystallite orientation. The refinement utilised (1) an exponential harmonic texture model; (2) spherical harmonics up to order 24; (3) a regular ODF sampling grid (\SI{5}{\degree} resolution); (4) multiple sample orientations as, during the scans, the samples were rotated about the azimuthal axis at a frequency of \SI{0.5}{Hz} to fully characterise the crystallite distribution. The reconstructed ODF maps show clear texture and are illustrated in Fig.~\ref{fig:ODFs}. This likely originates from the asymmetric nature of the orthorhombic unit cell. 

The texture is represented in multiples of random distribution (mrd), where mrd = 1 indicates random orientation, mrd $>1$ indicates preferred orientation, and mrd $<1$ indicates orientation depletion; typically mrd $>3$ or $<\frac{1}{3}$ are taken as evidence of a partially ordered or ordered solid. Our analysis revealed significant texture in the glycerol-dispersed sample with maximum mrd values exceeding 10 and minimum values below 0.01 in the inverse pole plot, and more than 1600 in the ODF, indicating a highly ordered crystalline arrangement within the glycerol matrix. This contrasts with the more uniformly distributed orientations in the pure powder sample.

The pole figures were generated for the principal crystallographic directions (normal direction [ND], rolling direction [RD], and transverse direction [TD]) to visualize the orientation distribution.   The texture analysis revealed that alanine crystals in glycerol adopt specific preferential orientations, likely due to interactions between the crystal faces and the glycerol matrix during sample preparation.
The refined texture model was incorporated into the final Rietveld refinement, which significantly improved the fit to the experimental data, confirming that the observed intensity variations were indeed due to preferred orientation rather than structural factors. These data are provided as supplemental data S1. 

\paragraph{Caption for Data S1.} 
\textbf{Raw and analysed crystallographic structures for Alanine} are provided as Data S1, including structural CIF files and textural fits and refinements used in this work. 

\subsection{Molecular Dynamics Simulations}
\label{SI:MolecularDynamics}
Molecular dynamics (MD) simulations were performed to investigate the interactions between alanine and glycerol molecules, with particular focus on hydrogen bonding characteristics. Rather than starting from a crystalline alanine structure, we deliberately employed a randomly positioned initial configuration for several methodologically sound reasons: (1) to establish an upper bound on the mixing thermodynamics between alanine and glycerol, (2) to avoid kinetic trapping issues associated with the prohibitively slow crystal dissolution process (which would exceed our accessible simulation timescale of tens of ns), (3) to directly probe the inherent molecular interactions without the confounding variable of crystal lattice energy, and (4) to model potential amorphous states of alanine that may be relevant in rapid cooling scenarios such as arise when a sample is placed into the DNP hyperpolariser. This approach allowed more efficient phase space sampling while providing a reference point for potential future crystalline-based simulations.

The simulation system was constructed using the Enhanced Monte Carlo (EMC) package (v4.1.5),\cite{InTVeld2003} a pre-processing module for LAMMPS.\cite{LAMMPS} A mixed system containing both alanine  and glycerol molecules was generated with a total of 1000 molecules at a temperature of 300 K and pressure of 1 atm. We elected to have a smaller number of molecules with periodic boundary conditions and let the system equilibrate for a longer period of time. The initial densities were set to $\SI{1.424}{g/cm^3}$ for alanine and $\SI{1.26}{g/cm^3}$ for glycerol. The PCFF (Polymer Consistent Force Field) was employed for all simulations, as implemented in the EMC and LAMMPS packages.\cite{Sun1998, Sun1998a} PCFF is a second-generation force field developed for organic molecules, polymers, and biomolecular systems, and has been extensively previously validated for simulating hydrogen bonding interactions in similar molecular systems, including alanine crystals.\cite{Meng2013, Sun2015, Yang2013a} The force field parameters were automatically generated by the EMC setup procedure. The integration timestep was set to 1 fs. Long-range electrostatic interactions were handled using the particle-particle particle-mesh (PPPM) method with a precision of 0.001. A cutoff of \SI{1}{nm} was applied for both van der Waals and Coulombic interactions. 

The simulation protocol consisted of the following steps:

\begin{itemize}
    \item System initialization: The mixed alanine-glycerol system was first constructed using EMC with random molecular placements based on the specified densities in two phases
\item Energy minimization: The system was energy-minimized to eliminate bad contacts using LAMMPS.
\item Equilibration: The system was equilibrated in two phases: 
\begin{itemize}
    \item Initial equilibration using a Langevin thermostat at \SI{300}{K} with a damping parameter of \SI{100}{fs}, combined with a distance-limited NVE integration (limit of \SI{0.1}{\angstrom}) for 1000 timesteps.
\item Production equilibration in the NPT ensemble at \SI{300}{K} and 1 atm pressure with temperature and pressure damping parameters of 100 fs and 1000 fs, respectively.
\end{itemize}
\item Production run: Following equilibration, a production run of 30 ns (30,000,000 timesteps) was performed in the NPT ensemble. The system state was recorded every 1000 timesteps (1 ps) for analysis. This took approximately 1 day. 
\end{itemize}

\begin{figure}
    \centering
    \includegraphics[width=\linewidth]{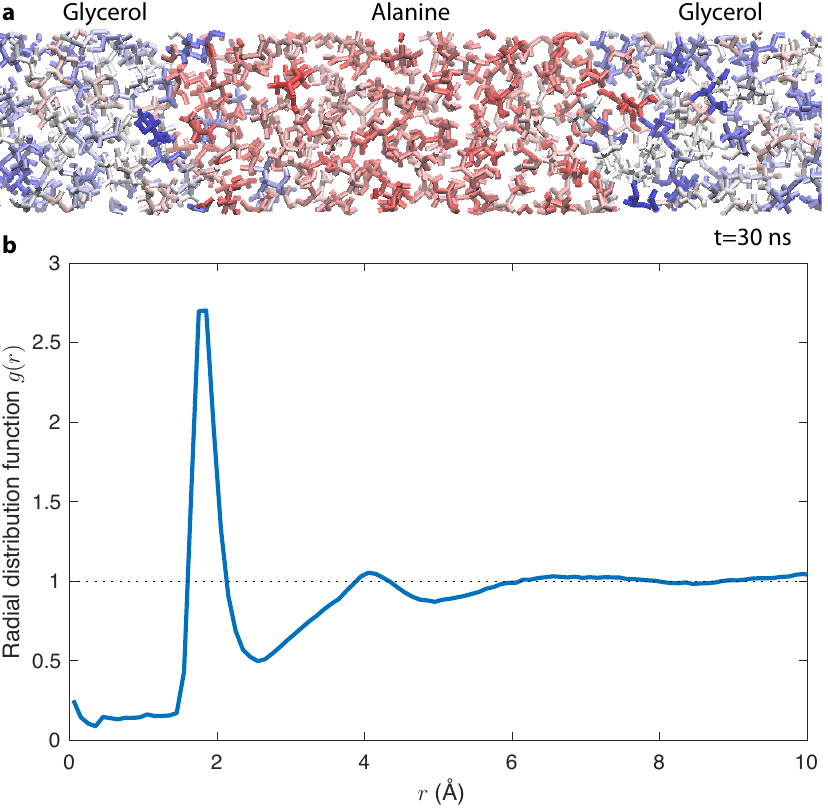}
    \caption{Molecular dynamic simulations of glycerol and alanine at \SI{300}{K}. \textbf{a}: The initial two-phase configuration remained approximately stable throughout the \SI{30}{ns} simulation, with limited penetration into molecular domains, indicating poor miscibility between the compounds; \textbf{b} the radial distribution function $g(r)$ for the system exhibits liquid-like characteristics with a prominent first coordination shell peak at approximately \SI{2}{\angstrom}, followed by diminishing oscillations approaching uniformity at larger distances, confirming the absence of long-range crystalline order.}
    \label{fig:md-simulations}
\end{figure}

Post-processing was performed in VMD,\cite{Humphrey1996} and density and the radial distribution function computed at the end of the equilibration process (c.f. Fig.~\ref{fig:md-simulations}). It was found that liquid glycerol did not penetrate substantially into the alanine domain on the timescale of the simulation, suggesting limited miscibility between the two substances under these conditions. The overall radial distribution function $g(r)$ exhibited features consistent with liquid-phase behaviour, including a prominent first coordination shell peak at approximately \SI{2}{\angstrom} followed by diminishing oscillations that approached uniformity at larger distances. This $g(r)$ profile confirms the absence of long-range crystalline order and indicates a liquid-like molecular arrangement with short-range spatial correlations characteristic of hydrogen-bonded systems.
While our simulation approach cannot definitively determine the global thermodynamic minimum of the system (which might involve a crystalline alanine phase separate from glycerol), it effectively establishes the relative stability of the mixed state and characterizes the molecular interactions at the interface between alanine and glycerol domains. The persistence of distinct domains throughout the \SI{30}{ns} simulation provides valuable information about the limited mutual solubility of these compounds under the studied conditions.

\subsection{Spin-dynamics simulations}
\label{si:spinach}

In an attempt to semi-empirically predict the asymmetric frequency sweep curve, a computational study on a model of the radical-damaged singular alanine crystal was undertaken. The spatial distribution of stable alanine radicals (SARs) was modelled using the known crystallographic data from the L-alanine structure provided by neutron diffraction. Since electron irradiation creates radicals at random positions within the crystal lattice, we employed bootstrap sampling (10,000 iterations) of possible radical positions to determine the statistical distribution of inter-radical distances, written using ASE in Python.\cite{HjorthLarsen2017} 

DNP simulations were performed using Spinach 2.4.5157 on the Arcus-C supercomputer with a magnetic field of 6.7~T. The basic single-electron, single-$^{13}$C system utilized the experimentally determined $g$-tensor of the stable alanine R1 radical with principal values $g=[2.0041, 2.0034, 2.0024]$ and eigenvectors from Sagstuen et al. Relaxation was modelled using the Weizmann formalism with electron $T_1 = 12.78$~ms, $T_2 = 0.3\, \mu$s, and nuclear $T_1 = 100$~s, $T_2 = 3$~ms at 1.4~K. 

For systems with multiple radical centres, simulations were extended to include two electron spins with slightly different $g$-tensors (differing by $10^{-5}$ in diagonal elements) and multiple $^{13}$C spins with coordinates derived from the crystal structure. We used a spherical tensor Liouvillian formalism. The swept microwave irradiation provided by our hardware was implemented as a frequency-modulated pulse sequence with the following parameters:
\begin{align}
\text{pulse-on duration} &= 9963~\mu\text{s} \\
\text{pulse-off duration} &= 48139~\mu\text{s} \\
\text{frequency sweep bandwidth} &= 25~\text{MHz} \\
\text{modulation rate} &= 1~\text{kHz}
\end{align}

The frequency sweep was divided into 25 discrete slices spanning from $-12.5$~MHz to $+12.5$~MHz relative to the centre frequency. For microwave sweep simulations, we examined 10 points between 187.5~GHz and 188.5~GHz. The build-up simulations were run to a maximum time of 60 seconds, corresponding to acquired experimental data.

The powder averaging was performed using the \texttt{icos-2ang-12pts} grid for the basic simulations and \texttt{rep-2ang-100pts-sph} for more detailed models. The polarization transfer was monitored using detection operators corresponding to the $^{13}$C $L_z$ Liouvillian terms. For extended simulations with multiple $^{13}$C nuclei, we incorporated 56 carbon positions from a supercell of the alanine unit cell (which contains four molecules), with the radical species positioned at sites 21 and 5 in the lattice. The electron-electron dipolar coupling was calculated from their spatial separation within the crystal structure. Periodic boundary conditions were included. The simulation ran on 200 nodes, each containing 48 core Cascade Lake processors (Intel Xeon Platinum 8268 CPU @ 2.90GHz) with 392 GB of ram, and took approximately 4 days to complete. 
\begin{figure}
    \centering
    \includegraphics[width=0.5\linewidth]{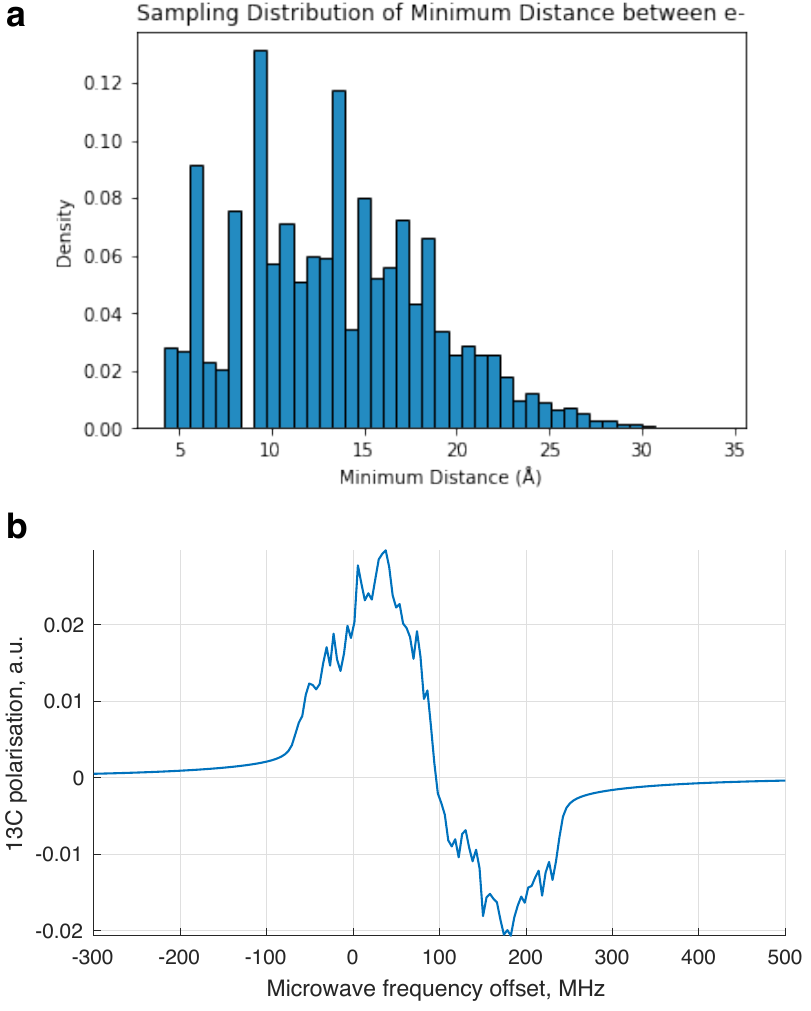}
    \caption{\textbf{a}: Simulated minimum electronic-radical distances based on a 70 kGy dose equivalent electron concentration utilising a stochastic damage algorithm in ASE. \textbf{b}: Predicted microwave sweep curve from Spinach using atomic coordinates from (\textbf{a}) with periodic boundary conditions and mimicking the experimental conditions under which DNP occurred.}
    \label{fig:Spinach}
\end{figure}

\subsection{Analytic models}
\label{si:analyticmodels}
Thermal mixing is a key mechanism in Dynamic Nuclear Polarization (DNP), widely taken to be dominant in samples used in clinical practice, and theoretically challenging. To quantitatively assess the DNP polarization behaviour, we implemented a numerical simulation based on Wenckebach's framework in the low temperature limit \cite{Wenckebach2017}.

In this model, the electron spin system is characterized by two parameters: the inverse electron Zeeman temperature $\alpha = \hbar/k_B T_Z$ and the inverse electron non-Zeeman temperature $\beta_{NZ} = \hbar/k_B T_{NZ}$. This assumption splits the electronic states into those that are polarised thermally (Zeeman) and those that will ultimately transfer energy to the nuclei in the irradiation phase of DNP (non-Zeeman). 
 This results in an electron spin polarization as a function of resonance frequency:

\begin{equation}
P_S(\omega) = \tanh\left[\frac{1}{2} \left(\omega_0\alpha + (\omega - \omega_0)\beta_{NZ}\right)\right]
\end{equation}

The evolution of the electron Zeeman and non-Zeeman temperatures is governed by a pair of coupled differential equations:

\begin{align}
F_1(\omega_m, \alpha, \beta_{NZ}) &= -2W(\omega_m) P_s(\omega_m,\alpha,\beta_{NZ}) - \frac{1}{T_1} \left[\int g(\omega) P_s(\omega,\alpha,\beta_{NZ}) \mathrm{d}\omega - P_L\right] \\ 
F_2(\omega_m, \alpha, \beta_{NZ}) &= -2W(\omega_m) (\omega_m - \omega_0 ) P_S(\omega_m, \alpha, \beta_{NZ}) - \frac{1}{T_1}\int g(\omega) (\omega_m - \omega_0) P_s(\omega,\alpha,\beta_{NZ})\,\mathrm{d}\omega 
\end{align}

\noindent where $g(\omega)$ is the normalised ESR spectral density with centre of mass $\omega_0$, $\omega_m$ is the applied microwave frequency, $W(\omega_m) = \frac{1}{2}\pi\omega_{1S}^2g(\omega_m)$ represents the microwave power, $\omega_{1S} = \gamma_S|B_1|$ is the electron Rabi frequency, $T_{1S}$ is the electron spin-lattice relaxation time, and $P_L = \tanh(\frac{1}{2}\omega_0\beta_L)$ is the thermal equilibrium polarization with $\beta_L = \hbar/k_BT_L$ being the inverse lattice temperature. High and low temperature limits are defined by the $\omega_0 \beta \approx 1$ boundary at roughly around \SI{1}{K}. 

We numerically solved this scheme in Mathematica by use of the explicit analytic calculation of its Jacobian. The function $g(\omega)$ was simulated by EasySpin under appropriate experimental conditions and a numerical interpolation function applied to extend its domain to $[0, \infty]$ (defining it as zero outside of the calculated region).

\begin{figure}
    \centering
    \includegraphics[width=0.8\linewidth]{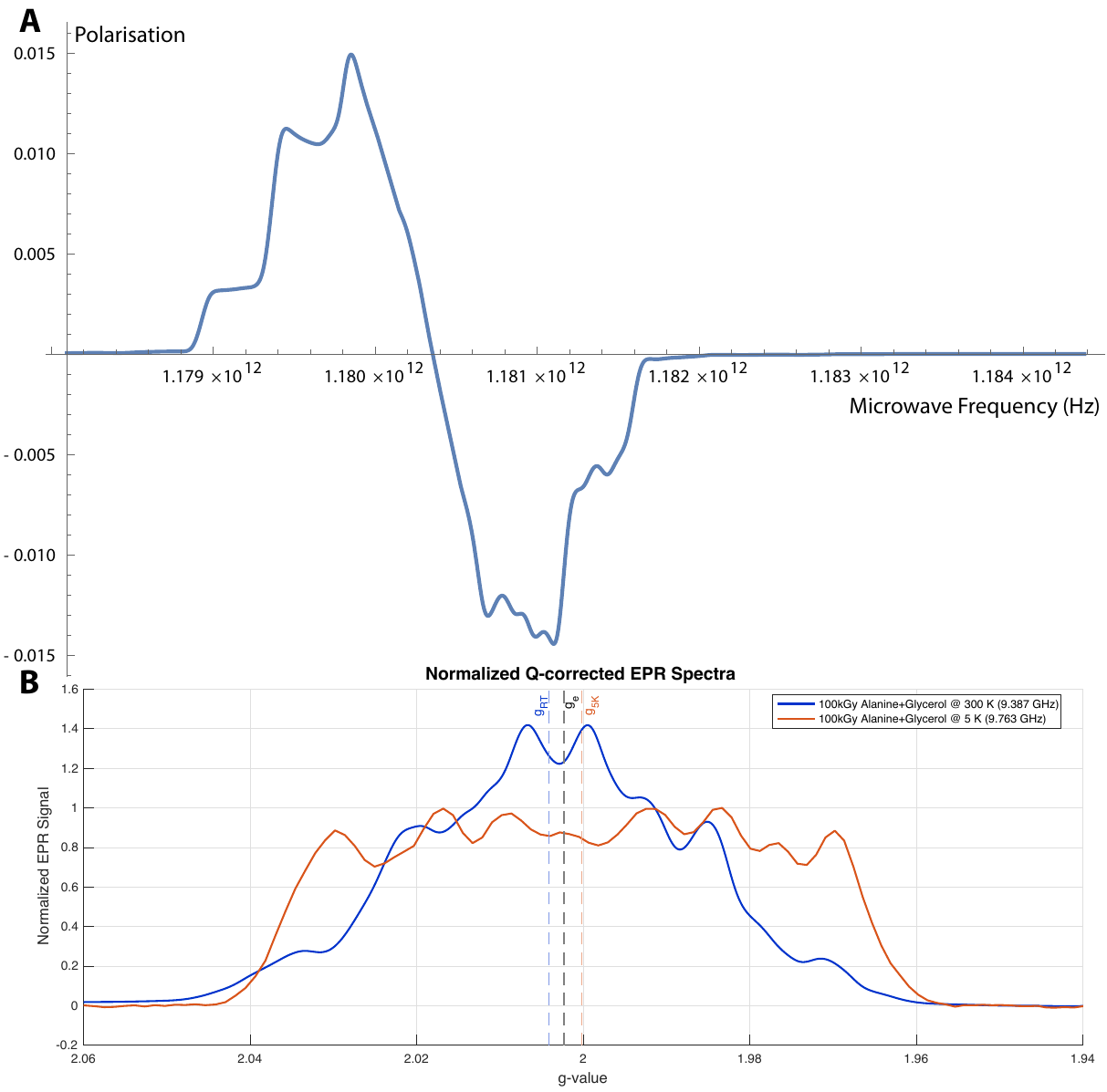}
    \caption{\textbf{A}: An analytic model of thermal mixing with the theoretical EPR lineshape given produces a bimodal build up curve. \textbf{B}: Integrated experimental low-temperature EPR Spectra obtained for the glycerol/alanine mixture compared to at room temperature (c.f. Fig.~\ref{fig:DNP}) indicated a shift in $g$. }
    \label{fig:thermalmixing}
\end{figure}

These equations were evaluated using adaptive Gaussian quadrature (\texttt{NIntegrate}) with specified domain constraints. To ensure numerical stability, the working precision was set to 20 digits, with recursion limits extended to handle the complex structure of the integrands. 
This process took approximately 1 day on a dual Xeon gold workstation with 768 GB ram, and the simulation results are shown in Fig.~\ref{fig:thermalmixing}\textbf{A}, with experimental integrated EPR lineshapes shown in Fig.~\ref{fig:thermalmixing}\textbf{B}. They produce a bimodal curve qualitatively similar to that in Spinach, and do not reflect experimental results.

\begin{figure}
    \centering
    \includegraphics[width=\linewidth]{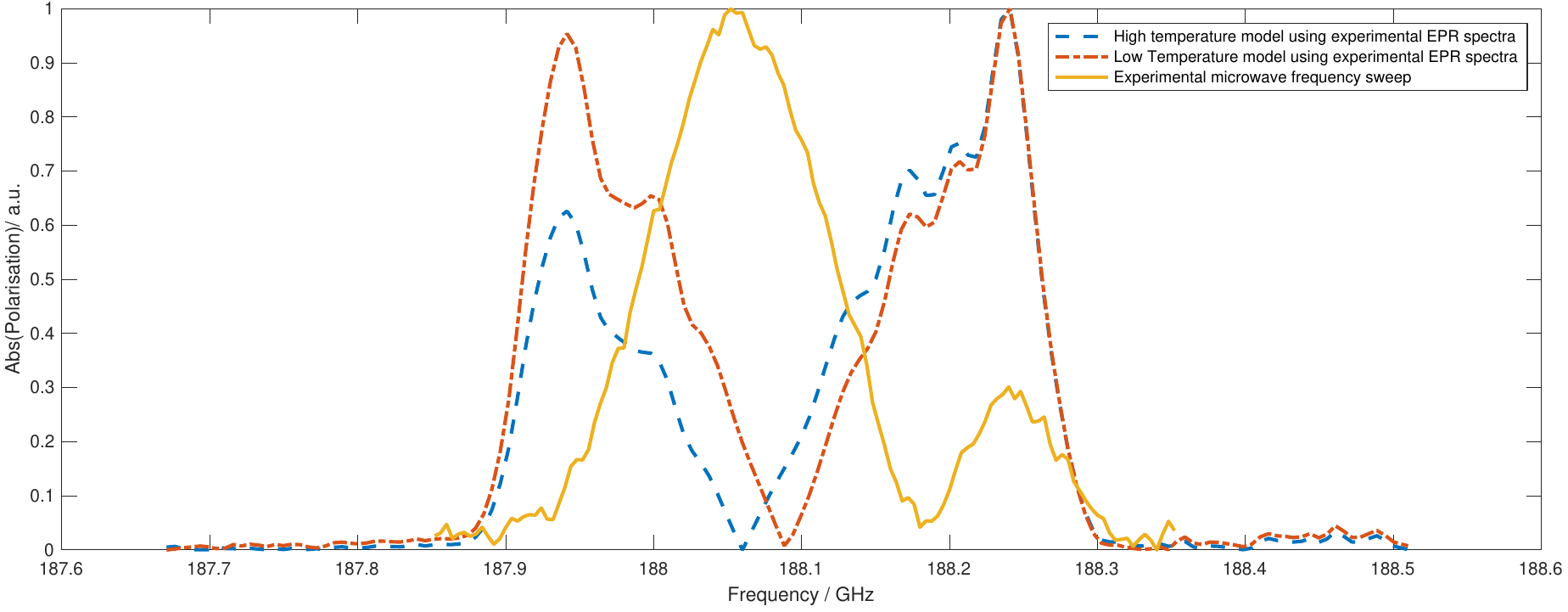}
    \caption{A semi-empirical model of the absolute value of the predicted enhancement curve of DNP as a function of frequency. The EPR spectrum used in this prediction was that obtained at \SI{5}{K} and lower field, scaled approximately for the system at hand. Whilst some degree of asymmetry is predicted by the Wenckebach model, it is not observed. An empirical shift of \SI{90}{MHz} aligns the curves, but does not resolve the predicted asymmetry. }
    \label{fig:ZoeModel}
\end{figure}

As a semi-empirical test with the observed (and not modellable) EPR lineshape obtained, a further simulation was conducted using a semi-empirically scaled EPR lineshape based on that obtained of the alanine/glycerol mixture at 5 K. This simulation was performed in Matlab (providing a novel reimplementation of the model above) and scaled the centre linearly and width (by approximately a factor of two) of the EPR line to be that appropriate at \SI{6.7}{T} under the assumption that these qualitative features would describe the system, as it is not possible for us to directly measure the EPR spectrum under DNP conditions of \SI{1.4}{K} and \SI{6.7}{K}. Both the high temperature and low temperature approximations as defined by Wenckebach have been solved separately and compared to experimental data. Even at \SI{30}{kGy} irradiation, where a small amount of bimodal behaviour is observed, this model (which accurately reproduces the behaviour of e.g. trityl radicals, TEMPO, Totapol, and other samples) does not fit the data obtained empirically, shown in Fig.~\ref{fig:ZoeModel}. We are confident that the bandwidth of the (commercially supplied) microwave excitation source exceeds that shown, and is not the limiting factor in these experiments. 

\subsection{QuantumESPRESSO}
\label{si:QE}

First-principles electronic structure calculations were performed using the plane-wave pseudopotential approach as implemented in the Quantum ESPRESSO (QE) package (version 7.2) \cite{Giannozzi2009,Giannozzi2017}. The Perdew-Burke-Ernzerhof (PBE) generalized gradient approximation \cite{Perdew1996} was used for the exchange-correlation functional. Projector augmented-wave (PAW) pseudopotentials were employed for carbon and oxygen atoms, while ultrasoft pseudopotentials were used for hydrogen and nitrogen atoms, all obtained from the Standard Solid-State Pseudopotential (SSSP) library (efficiency version) \cite{Prandini2018}.

The crystal structure of L-alanine was based on the available neutron diffraction data, with a unit cell containing 52 atoms (four alanine molecules). The orthorhombic unit cell parameters were $a = 5.7880$ \AA, $b = 6.0360$ \AA, and $c = 12.3420$ \AA, with space group P2$_1$2$_1$2$_1$ (No. 19).

Calculations were performed with the following parameters: kinetic energy cutoff for wavefunctions of 60 Ry; kinetic energy cutoff for charge density of 480 Ry; 100 Kohn-Sham bands computed to ensure proper description of the conduction bands; electronic convergence threshold of $1.0 \times 10^{-6}$ Ry; mixing parameter (beta) of 0.6.

\subsubsection*{Brillouin Zone Sampling}

For the band structure calculations, a path along high-symmetry points in the first Brillouin zone was sampled. A total of 392 k-points were used along the path, which included the following segments: $\Gamma \rightarrow$ X; X $\rightarrow$ S; S $\rightarrow$ Y; Y $\rightarrow \Gamma$; $\Gamma \rightarrow$ Z; Z $\rightarrow$ U; U $\rightarrow$ R; R $\rightarrow$ T.

The k-point mesh was generated in crystal coordinates, with a dense sampling to ensure smooth band dispersion.

\subsubsection*{Computational Resources}

The calculations were performed on a parallel computing architecture using 40 MPI processes distributed across a single node and using nearly 1 TB of ram. The electronic structure calculation for the system consumed approximately 24 hours of CPU time and 31 hours of wall time, highlighting the computational intensity of modelling this organic crystal system.

\subsubsection*{Analysis of Electronic Structure}

\begin{figure}
    \centering
    \includegraphics[width=\linewidth]{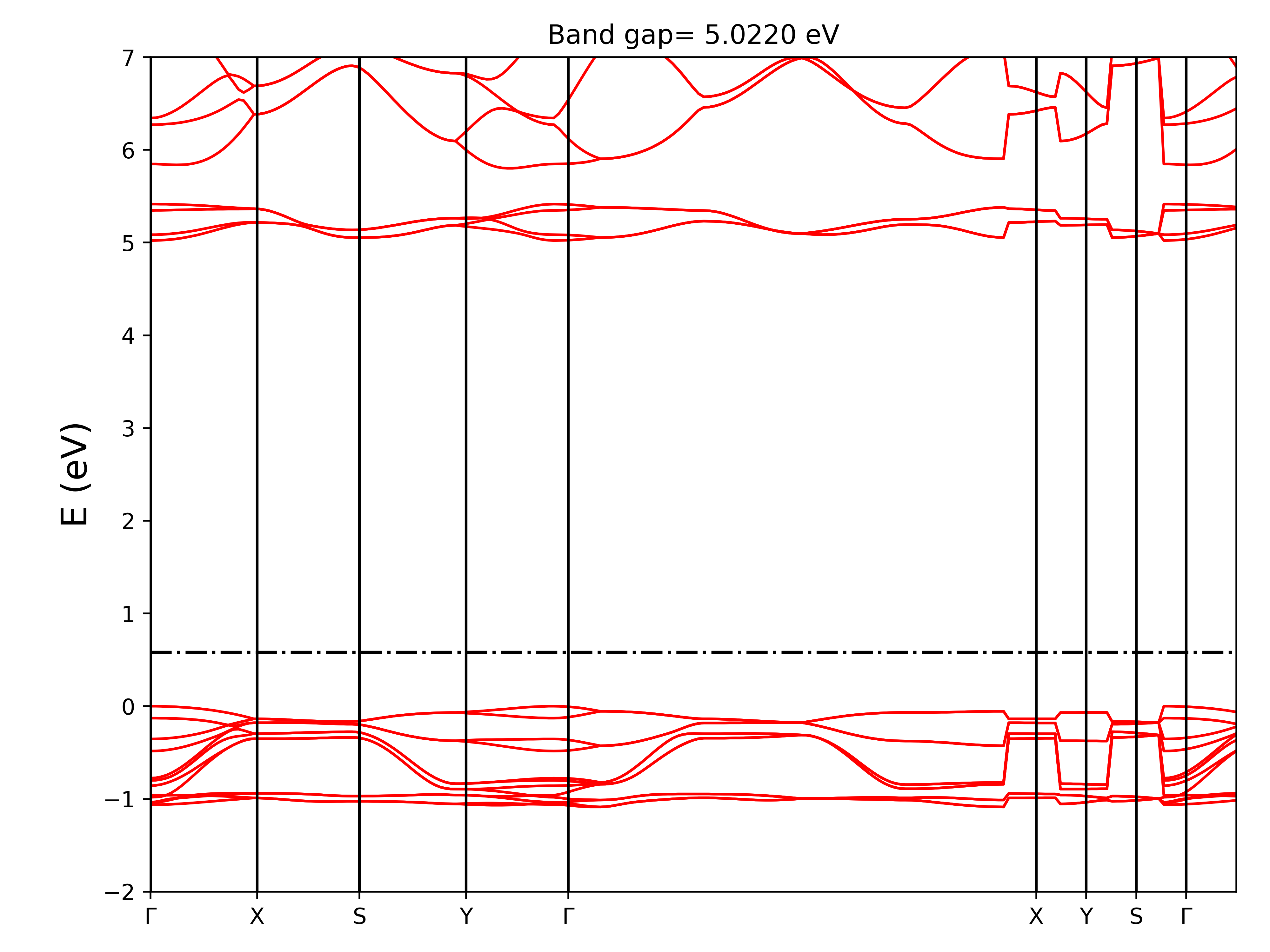}
    \caption{The estimated bandstructure of alanine crystals predicts that it is a relatively large bandgap insulator.}
    \label{fig:QE}
\end{figure}

The band structure calculations revealed a fundamental electronic band gap of approximately 5.02 eV (calculated as the difference between the highest occupied state at 0.58 eV and the lowest unoccupied state at 5.60 eV). The energy eigenvalues were extracted from the full band structure calculation and used to generate the band dispersion plots shown in Fig.~\ref{fig:QE}. This compares favourably to one reported experimental values of \SI{5.4}{eV},\cite{Suresh2020} especially as DFT studies using PBE functionals tend to underestimate band gaps by 0.5-1.0 eV. 

The total energy of the system was $-668.849$ Ry, with contributions from: one-electron term ($-362.629$ Ry); Hartree term (237.420 Ry); exchange-correlation term ($-157.484$ Ry); Ewald term ($-236.071$ Ry); one-center PAW contribution ($-150.084$ Ry).

The relatively large bandgap supports the stability of radical environments, and rules out one (admittedly outlandish) potential mechanism for DNP: that of the well-resolved solid effect.

\end{document}